\documentclass[preprint,12pt,authoryear]{elsarticle}

\usepackage{graphicx}
\usepackage{amssymb, amsfonts, amsthm, mathtools}
\usepackage{latexsym, hyperref, url, moreverb, color}
\usepackage{enumerate}
\usepackage{xspace}
\usepackage[margin=1in]{geometry}
\usepackage{xcolor}
\usepackage[utf8]{inputenc}
\usepackage[T1]{fontenc}

\graphicspath{{figs/}}
\renewcommand{\d}{\text{d}}
\newcommand{\Exp}{\mathbb{E}}
\newcommand{\Entropy}{H}
\newcommand{\Multinom}{\mathcal{M}}
\newcommand{\Normal}{\mathcal{N}}
\newcommand{\Real}{\mathbb{R}}
\newcommand{\A}{\boldsymbol{A}}
\newcommand{\B}{\boldsymbol{B}}
\newcommand{\I}{\boldsymbol{I}}
\newcommand{\D}{\boldsymbol{D}}
\newcommand{\V}{\boldsymbol{V}}
\newcommand{\Yspace}{\mathcal{Y}}

\newcommand{\x}{\boldsymbol{x}}

\renewcommand{\u}{\boldsymbol{u}}
\newcommand{\mub}{\boldsymbol{\mu}}
\newcommand{\Sigmab}{\boldsymbol{\Sigma}}
\newcommand{\pib}{\boldsymbol{\pi}}
\newcommand{\thetab}{\boldsymbol{\theta}}
\renewcommand{\hat}[1]{\widehat{#1}}
\newcommand{\T}{{}^{\!\top}}

\DeclareMathOperator*{\argmax}{argmax\xspace}

\newcommand{\BIC}{\mathrm{BIC}}

\newcommand{\pkg}[1]{\texttt{#1}}
\newcommand{\EntropyGMM}{\textsf{EntropyGMM}}
\newcommand{\EntropyGMMb}{\textsf{EntropyGMMb}}
\newcommand{\KLnn}{\textsf{KLnn}}
\newcommand{\SR}[2]{\textcolor{black}{#2}}

\begin{document}

\begin{frontmatter}



\title{Mixture-based estimation of entropy}


  \author[1]{St\'ephane Robin}
  \affiliation[1]{organization={MIA-Paris, Université Paris-Saclay -- AgroParisTech -- INRAE},
            city={Paris},
            postcode={75005}, 
            country={France}}
            
  \author[2]{Luca Scrucca}
  \affiliation[2]{organization={Department of Economics, Università degli Studi di Perugia},
            city={Perugia},
            postcode={06123}, 
            country={Italy}}

\begin{abstract}
The entropy is a measure of uncertainty that plays a central role in information theory. When the distribution of the data is unknown, an estimate of the entropy needs be obtained from the data sample itself. We propose a semi-parametric estimate, based on a mixture model approximation of the distribution of interest. The estimate can rely on any type of mixture, but we focus on Gaussian mixture model to demonstrate its accuracy and versatility. 
Performance of the proposed approach is assessed through a series of simulation studies. 
We also illustrate its use on two real-life data examples.
\end{abstract}




\end{frontmatter}


\section{Introduction} \label{sec:intro}
\emph{Differential entropy} for continuous random variables, also called simply the \emph{entropy}, is an extension of the concept of entropy introduced by \citet{Shannon:1948}. 
Consider a multivariate continuous random variable $Y \in \Real^p$ with probability density function $f(y)$. The entropy of $Y$ is defined as
\begin{equation} \label{eq:entropy}
\Entropy(Y) = - \int_{\Yspace} f(y) \log f(y) \d y = - \Exp [\log f(Y)],
\end{equation}
where $\Yspace = \{y : f(y) > 0 \}$ is the support of the random variable.
The entropy is a measure of the average uncertainty or information content in a random variable, and forms one of the core ideas in \emph{Information Theory} \citep{Shannon:1948}. For a comprehensive introduction see \citet{Cover:Thomas:2006}. 

Estimation of the entropy in \eqref{eq:entropy} can be pursued in different ways. A general approach that can always be applied, either in the univariate and multivariate case, being the random variable discrete or continuous, is Monte Carlo (MC) integration. This technique directly approximates equation \eqref{eq:entropy} by drawing iid samples $\{\widetilde{y}_i\}_{i=1}^s$ of size $s$ from $f(y)$ and then compute:
\begin{equation*}
\Entropy_{\text{MC}}(Y) = -\frac{1}{s} \sum_{i = 1}^s \log f(\widetilde{y}_i).
\end{equation*}
By the law of large numbers, $\Entropy_{\text{MC}}(Y) \to \Entropy(Y)$ as $s \to \infty$, thus MC approximation guarantees convergence to the true value of the entropy, although a very large MC sample size is required to reasonably approximate the expected value.

A more efficient approach is available if we are willing to assume a specific distributional form $f(y) = f(y; \thetab)$, which depends on the unknown (possibly a vector) parameter $\thetab$. 
Closed-form expressions for the differential entropy of several univariate distributions are readily available in the literature. For a review see \citet{Michalowicz:etal:2014}. 
However, only few multivariate distributions admit an analytical solution, most of them being linked to the multivariate Gaussian distribution: see \citet[pp. 172--173]{Michalowicz:etal:2014} for the multivariate Gaussian, and \citet{MSL14} or \citet{QSL16} for related distributions.
{In particular, if} $Y$ is a multivariate Gaussian random variable, i.e. $Y \sim \Normal(\mub, \Sigmab)$, where $\mub$ is the mean vector and $\Sigmab$ the covariance matrix, then the entropy can be shown to be given by
\begin{equation} \label{eq:entGauss}
\Entropy(Y) = \frac{1}{2} \log ( (2\pi e)^p |\Sigmab| ).
\end{equation}
In the univariate case the above formula simplifies to $\Entropy(Y) = \dfrac{1}{2} \log (2\pi e \sigma^2)$.
Thus, when a closed-form expression is available the problem becomes that of estimating the unknown parameter $\thetab$. If $\hat{\thetab}$ is the maximum likelihood estimate of $\thetab$, then the estimate of the entropy is guaranteed to be both asymptotically unbiased and efficient \citep[see][Theorem 7.3, p. 183]{Kay:1993}.

If we cannot assume a particular known distribution, we need to estimate $f(y)$. Simple nonparametric estimators are the histogram, where the range of observed values are divided into discrete ``bins'', and kernel density estimation. Both estimators depend on the size of the dataset and on the choice of tuning parameters, such as the bin width or the kernel bandwidth. 
The R package \pkg{infotheo} \citep{pkg:infotheo} provides estimates of Information-Theory measures, including the entropy, for both univariate and multivariate variables based on histogram estimators.
Two nonparametric procedures for estimating the Mutual Information (see \ref{sec:mi}), and so the entropy as by-product, are provided by the R package \pkg{mpmi} \citep{pkg:mpmi}, using a kernel density estimator with and without bias-correction \citep{Pardy_etal:2018}, and by the R package \pkg{rmi} \citep{pkg:rmi}, using a local nearest neighbor estimator based on Gaussian kernel and kNN selected bandwidth \citep{Gao:etal:2017}.

A semiparametric estimate of $f(y)$ can also be obtained using mixture models, where the unknown density is approximated by a convex linear combination of one or more probability density functions. According to \citet[p.~237]{FruhwirthSchnatter:2006}, 
"[f]inite mixture distributions can be used to derive arbitrarily accurate approximations to practically any given probability distribution, provided that the number of components is not limited". 
{Following this line of thought,} we propose a generic approach to estimate the entropy of an arbitrary (multivariate) distribution $f$ using a finite mixture as an approximation for $f$. We show that a natural estimate can be derived as a side product of the celebrated EM algorithm \citep{DLR77}. Because of their popularity, we focus here on Gaussian mixture models (GMMs), but the approach can be extended to any mixture.

In this contribution we propose a method for estimating the entropy of a (possibly multivariate) distribution, using a finite mixture as an approximation of the distribution of interest $f$. 
The paper is organized as follows. 
Section~\ref{sec:entropy_fmm} derives the entropy for a general mixture, while Section~\ref{sec:estimation} contains the proposed estimation method. 
Section~\ref{sec:sim} presents some simulation studies to evaluate the performance of the proposed entropy estimation procedure and compares it with some of the estimation methods mentioned above.
Section~\ref{sec:app} provides examples of real data applications, namely image quantization and segmentation (\ref{sec:image}), and structure identification of tree-shaped graphical models (\ref{sec:tree}). 
The final section provides some concluding remarks.


\section{Entropy of finite mixture distributions} \label{sec:entropy_fmm}
Suppose that the density of $Y$ can be expressed as a finite mixture of the form
\begin{equation} 
\label{eq:f}
f(y) = \sum_{k=1}^K \pi_k \psi_k(y),
\end{equation}
where $\pi_1, \ldots, \pi_K$ are the mixing weights, which satisfy $\pi_k > 0$ and $\sum_{k=1}^K \pi_k = 1$, and $\psi_k(\cdot)$ for $k=1,\ldots,K$, are the component densities of the mixture. Usually, $\psi_k(y) = \psi(y; \thetab_k)$, i.e.\ component densities share the same distribution, but with different parameter $\thetab_k$. In the sequel, we use $\psi(\cdot; \thetab)$ as a notation for a generic (parametric) distribution, and use $\phi(\cdot; \mub, \Sigmab)$ specifically for the Gaussian distribution.

The mixture distribution in \eqref{eq:f} can be equivalently described by the following hierarchical generative model:
\begin{equation} \label{eq:mixt}
Z \sim \Multinom(1; \pib), \qquad
Y \mid (Z = k) \sim \psi_k(y),
\end{equation}
where $\Multinom(1; \pib)$ is the multinomial distribution with $\pib = (\pi_1, \pi_2, \ldots, \pi_K)$.
From the definition of entropy in \eqref{eq:entropy}, and considering the hierarchical model in \eqref{eq:mixt}, we have that
\begin{align} 
\label{eq:H}
\Entropy(Y) 
& = - \Exp_Z \left[ \Exp_{Y \mid Z} \left[ \log f(Y) \right]\right]
  = - \sum_{k=1}^K \pi_k \Exp\left[ \log f(Y) \mid Z=k \right] \nonumber \\
& = - \sum_{k=1}^K \pi_k \int_{\Yspace} \psi_k(y) \log f(y) \d y.
\end{align}
Then, the decomposition of $\log f(y)$ that inspires the EM algorithm states that
\begin{align*}
\log f(y) 
& = \Exp \left[ \log p(Z, Y) \mid Y = y \right] - \Exp \left[ \log p(Z \mid Y) \mid Y = y \right] \\
& = \sum_{k=1}^K \tau_k(y) \left( \log \pi_k + \log \psi_k(y) - \log \tau_k(y) \right),
\end{align*}
where $\tau_k(y)$ indicates the conditional probability{, that is }
$$
\tau_k(y) = \Pr\left(Z = k \mid Y = y\right) 
          = \frac{\pi_k \psi_k(y)}{\sum_{g=1}^K \pi_g \psi_g(y)}.
$$
A regular EM algorithm provides estimates of the quantities $\pi_k$, $\psi_k(y)$, and thus $\tau_k(y)$. 
However, according to equation \eqref{eq:H}, we still need to evaluate the integral
\begin{align} \label{eq:Ak}
A_k = \int_{\Yspace} \psi_k(y) \log f(y) \d y. 
\end{align}

\paragraph{Gaussian mixtures} GMMs correspond to the case where
the general mixture in \eqref{eq:f} can be written as
\begin{equation}
\label{eq:gmm}
f(y) = \sum_{k=1}^K \pi_k \phi( y ; \mub_k, \Sigmab_k),
\end{equation}
where $\{\pi_1, \pi_2, \dots , \pi_{K-1}, \mub_1, \ldots,  \mub_K, \Sigmab_1, \dots, \Sigmab_K \}$ are the parameters of the mixture model, with $(\pi_1,\pi_2, \dots, \pi_K)$ the mixing weights, and $\phi(y ; \mub_k, \Sigmab_k)$ the underlying multivariate Gaussian density function of \textit{k}th component with  mean vector $\mub_k$ and covariance matrix $\Sigmab_k$. No closed-form expression exist for the entropy of such a mixture, but several approximations of it have been proposed in the literature \citep{Julier:Uhlmann:1996, Goldberger:Aronowitz:2005, Hershey:Olsen:2007, Huber:etal:2008}, which are recalled in \ref{sec:approxEntropyGMM}.  


\section{Estimation} \label{sec:estimation}
We now describe how to use the formulas derived {in the previous} section 
to get a mixture-based estimate of the entropy of a distribution $f$. We then turn to the specific use of Gaussian mixture models and discuss how the proposed estimate can be implemented in practice.

\subsection{Estimating the entropy}

Suppose we have observed an iid sample $\{y_i\}_{i=1}^n$ from the distribution $f$. An EM algorithm can be used to fit a finite mixture model to the sample $\{y_i\}_{i=1}^n$, yielding estimates for the mixing weights $\hat{\pi}_k$, the component densities $\hat{\psi}_k(y_i)$, the conditional probabilities $\hat{\tau}_k(y_i)$, and a plug-in estimate of the density for each data point 
$$
\hat{f}(y_i) := \sum_{k=1}^K \hat{\pi}_k \hat{\psi}_k(y_i). 
$$

An estimate of $A_k$, defined in \eqref{eq:Ak}, is still needed, and this can be obtained by a change of measure. 
Indeed, one may observe that 
\begin{align*}
 A_k = \int f(y) \frac{\psi_k(y)}{f(y)} \log f(y) \d y,
\end{align*}
which is well defined because, from the definition of the mixture in \eqref{eq:f}, $f$ dominates each $\psi_k$.
Now because the $y_i$ are iid from $f(y)$, we may define
\begin{align*}
\hat{A}_k = \sum_{i=1}^n w_k(y_i) \log \hat{f}(y_i),
\end{align*}
where the weights are defined as
$$
w_k(y_i) 
= \frac{\hat{\psi}_k(y_i)}{\hat{f}(y_i)} 
\propto \frac{\hat{\tau}_k(y_i)}{n\hat{\pi}_k}, 
\qquad\text{with }
\sum_{i=1}^n w_k(y_i) = 1.
$$
Finally, we end up with the following \emph{mixture-based estimate of the entropy}:
\begin{align}
\label{eq:est}
\hat{\Entropy}(Y) 
& = - \sum_{k=1}^K \hat{\pi}_k \hat{A}_k
= - \sum_{k=1}^K \hat{\pi}_k \left( \sum_{i=1}^n \frac{\hat{\tau}_k(y_i)}{n\hat{\pi}_k} \log \hat{f}(y_i) \right)
\nonumber \\
& = - \frac{1}{n} \sum_{i=1}^n  \sum_{k=1}^K  \hat{\tau}_k(y_i) \log \hat{f}(y_i) 
  = -\frac1n \sum_i \log \hat{f}(y_i),
\end{align}
where the last equality comes from noting that $\sum_{k=1}^K  \hat{\tau}_k(y_i) = 1$ for any $y_i$.

%

The estimator in equation~\eqref{eq:est} can also be obtained from a different perspective following a plug-in principle. 
Given an iid sample $\{y_i\}_{i=1}^n$ we could estimate $\Entropy(Y)$ by replacing the theoretical expectation with its empirical counterpart, i.e.
$$
\hat{\Entropy}(Y) = -\hat{\Exp}_Y [\log \hat{f}(Y)] = - \frac{1}{n} \sum_{i=1}^n \log \hat{f}(y_i).
$$
Therefore, the proposed estimator is semi-parametric in the sense that the density $\hat{f}$ is estimated parametrically using a mixture distribution, whereas the estimate of the expectation $\hat{\Exp}_Y$ is \SR{}{based on the empirical distribution and is therefore} nonparametric. 
The final estimator is simple and intuitive, provided that a mixture-based estimator of the log-density is available and it is evaluated on the sample data points used by the EM algorithm for model fitting. 
Furthermore, despite its similarity to the MC estimator, a remarkable advantage of the proposed estimator is that it does not require any Monte Carlo sampling, apart from the sample data themselves.

\subsection{GMM-based entropy estimate}

The estimator we propose is based on a GMM approximation of the true density $f$. Indeed, GMMs can approximate any continuous density with arbitrary accuracy provided the model has a sufficient number of components and the parameters of the model are correctly estimated \citep{Escobar:West:1995, Roeder:Wasserman:1997}.

A first way to estimate $H(Y)$ would consist in fitting a GMM --~as defined in equation \eqref{eq:gmm}~-- to the observed sample $\{y_i\}_{i=1}^n$ and then to plug the estimates $\{\hat\pi_k\}_{k=1}^K$, $\{\hat\mub_k\}_{k=1}^K$, and $\{\hat\Sigmab_k\}_{k=1}^K$ into one of the approximations of $H(Y)$ given in \ref{sec:approxEntropyGMM}. Such an estimate would suffer from both the possibly poor accuracy of the approximation of the theoretical entropy of a Gaussian mixture and from the uncertainty of the parameter estimates.

We rather opt for the semi-parametric estimator $\hat H(Y)$ defined in equation \eqref{eq:est}, setting
$$
\hat f(y):= \sum_{k=1}^K \hat\pi_k \phi(y; \hat\mub_k, \hat\Sigmab_k).
$$
To our knowledge, our estimate is the only one that is directly connected to the estimation procedure.

\subsection{Practical implementation}

The estimation procedure we propose relies on a mixture-based density estimate, which needs to be precisely defined. In case of GMMs both the number of components $K$ and the form of the covariance matrix within each component must be selected.
Parsimonious parametrizations of covariance matrices for GMMs can be obtained by adopting the eigen-decomposition $\Sigmab_k = \lambda_k\V_k\D_k\V\T_k$ \citep{Banfield:Raftery:1993, Celeux:Govaert:1995}, where $\lambda_k$ is a scalar controlling the volume of the ellipsoid, $\D_k$ is a diagonal matrix controlling its shape, and $\V_k$ is an orthogonal matrix controlling the orientation of the ellipsoid.

Unlike many applications, such those involving clustering of data, the analysis of these features is not our main interest here, since we do not aim to obtain a precise description of the components distribution.
Still, a wide variety of models can be considered and a classical option consists in choosing the 'best one'. The Bayesian information criterion \citep[BIC;][]{Schwarz:1978} is likely the most popular for GMMs. For a given model $m$, it is defined as $\BIC(m) = 2\hat{\ell}_m - \nu_m \log(n)$, where $\hat{\ell}_m$ stands for the maximized log-likelihood of the data sample of size $n$ under model $m$, and $\nu_m$ for the number of independent parameters to be estimated. 
For GMMs the latter depends on the number of mixture components and the assumed within-component covariance matrices. 
Among the set of $M$ models considered, the one chosen is then
$$
\hat{m} = \argmax_{m = 1,\dots, M} \; \BIC(m).
$$
Some results about the consistency of BIC are available under the assumption that the likelihood is bounded \citep{Keribin:2000} and for selecting the number of mixture components in the case of univariate density estimation \citep{Roeder:Wasserman:1997}. 

Finally, we should mention that, because entropy estimation can be considered for many different purposes, the GMM that is actually used for this estimation can be selected in a completely problem-specific manner, as discussed in Sections \ref{sec:chisq:dist}--\ref{sec:mi}, and illustrated in the applications in Section \eqref{sec:tree}.


\section{Simulation studies} \label{sec:sim}
In this section we present a series of simulation studies conducted to assess the accuracy of the proposed entropy estimate for several distributions with closed-form expression for the entropy.
Furthermore, we provide a comparison with some common estimation methods implemented in R packages available on CRAN (\url{https://cran.r-project.org}).
In the following list, the methods under comparison and the corresponding labels used in Figures~\ref{fig:bivgauss}--\ref{fig:ChiSquared} are described.
Further details are provided in Section~\ref{sec:intro}.

\begin{itemize}

\item{\sf Entropy[MLE]}: entropy computed from closed-form expression with MLEs plugged-in for unknown parameters;

\item{\sf EntropyGMM}: entropy estimator based on our proposal in equation \eqref{eq:est} for GMMs;

\item{\sf UT, VAR, SOTE}: approximated entropy estimators for GMMs described in \ref{sec:approxEntropyGMM} and available in the package \pkg{ppgmmga} \citep{pkg:ppgmmga};

\item{\sf infotheo$|$*$|$**}: histogram-based estimation of the entropy using discretization method \textsf{*} (\textsf{eqfreq} = equal frequencies for each variable; \textsf{eqwidth} = equal width binning algorithm for each variable; \textsf{gleqwidth} = global equal width over the range of all the variables) and estimation method \textsf{**} (\textsf{emp} = estimator based on the empirical probability distribution; \textsf{mm} = Miller-Madow asymptotic bias corrected empirical estimator; \textsf{shrink} = shrinkage estimator based on the entropy of the Dirichlet probability distribution; \textsf{sg} = Schurmann-Grassberger estimator based on the entropy of a Dirichlet probability distribution). These estimators are available in the package \pkg{infotheo} \citep{pkg:infotheo};

\item{\sf mpmi$|$*}: kernel density-based estimator with \textsf{*} indicating without (\textsf{mi}) or with (\textsf{bcmi}) bias correction, and available in the package \pkg{mpmi} \citep{pkg:mpmi};

\item{\sf lnn}: local nearest neighbour-based estimator available in the package \pkg{rmi} \citep{pkg:rmi};

\item{\sf KLnn}: \citet{KL87} nearest neighbour estimator available in the package \pkg{IndepTest} \citep{pkg:indeptest}.
\end{itemize}

\subsection{Mixed-Gaussian distribution}
\label{sec:mixedgauss:dist}

\begin{figure}[htb]
\centering
\includegraphics[width=0.8\textwidth]{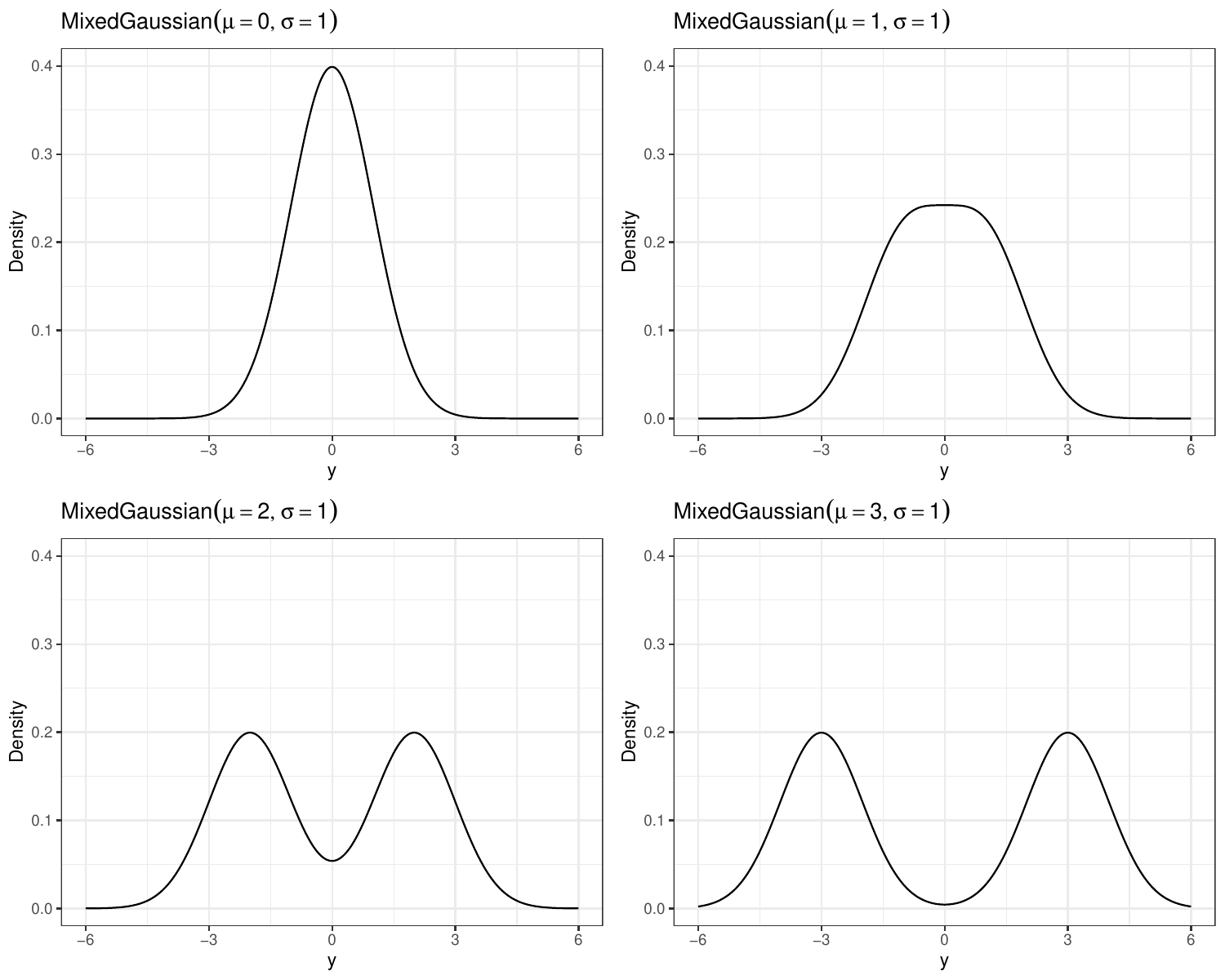}
\caption{\small Density plots of selected mixed-Gaussian distributions.}
\label{fig1:mixedGauss}
\end{figure}

The mixed-Gaussian distribution is often considered as a noise model in a number of signal processing applications \citep{Wang:Wu:2007}. It can be defined by splitting a Gaussian distribution $\Normal(0, \sigma^2)$ into two parts, centring one half at $+\mu$, and the other at $-\mu$ and summing the resultants. The density function is thus given by
$$
f(y) = \frac{1}{2\sqrt{2\pi}\sigma} \left( \exp\{-(x-\mu)^2/2\sigma^2\} + \exp\{-(x+\mu)^2/2\sigma^2\} \right)
\qquad -\infty < y < \infty,
$$
and has mean zero and variance equal to $\mu^2 + \sigma^2$. 
Clearly, the mixed-Gaussian distribution is a particular form of Gaussian mixture with $K=2$ components having equal proportions, means $(-\mu, \mu)$ and common variance $\sigma^2$.
Figure~\ref{fig1:mixedGauss} shows some shapes obtained by changing the parameter $\mu$. As $\mu$ increases from $0$, in which case is equivalent to the standard Gaussian distribution, the distribution increasingly shows a bimodal structure. 

\citet{Michalowicz:etal:2008} derived the entropy for the mixed-Gaussian distribution, which can be computed as
\begin{equation*}
\Entropy(Y) = \frac{1}{2} \log(2 \pi e \sigma^2) + (\alpha^2 - I),
\label{eq:entmixedgauss}
\end{equation*}
where $(\alpha^2 - I)$ is a function of $\alpha = \mu/\sigma$ and it is tabulated in \citet[][Table~1]{Michalowicz:etal:2008}. The fact that a closed-form solution is available warranties its inclusion in the simulation study.

\begin{figure}[htb]
\centering
\includegraphics[width=\textwidth]{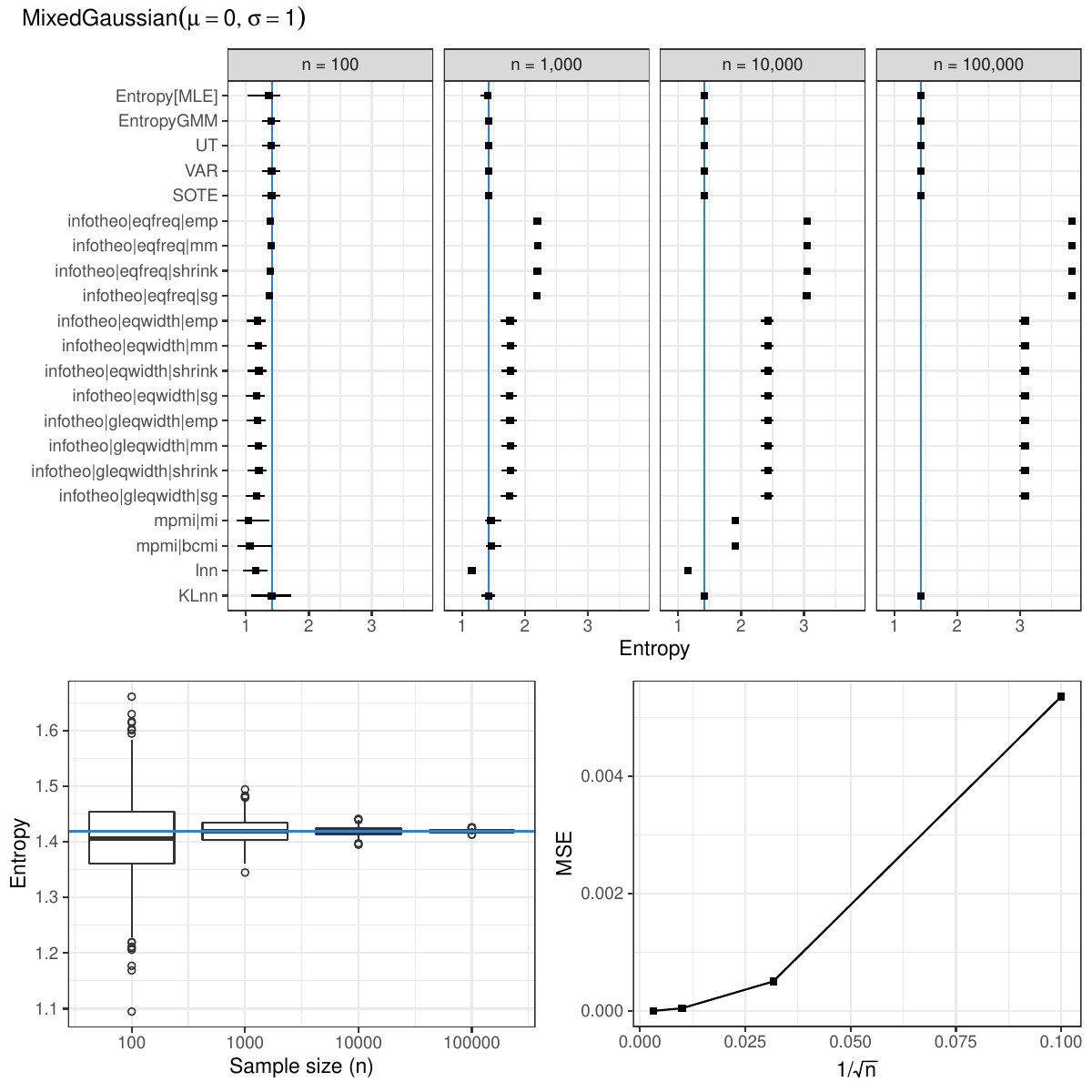}
\caption{\small Simulation results for univariate data generated from MixedGaussian($\mu= 0, \sigma = 1)$. The top panel shows dotcharts of entropy estimates for increasing sample sizes; each facet contains for each method under comparison the average entropy (square symbol) and the $(2.5\%,97.5\%)$-percentiles (horizontal lines) from 1000 simulation runs; blue vertical lines represent the theoretical entropy value. Bottom panels show the simulation results for the mixture-based entropy estimator. The bottom-left panel reports the box-plots for increasing sample sizes, with horizontal blue line representing the theoretical value of the entropy. The bottom-right panel reports the mean squared error (MSE) as function of the inverse-square root of sample sizes for checking $\sqrt{n}$-consistency.}
\label{fig2:mixedGauss}
\end{figure}

\begin{figure}[htb]
\centering
\includegraphics[width=\textwidth]{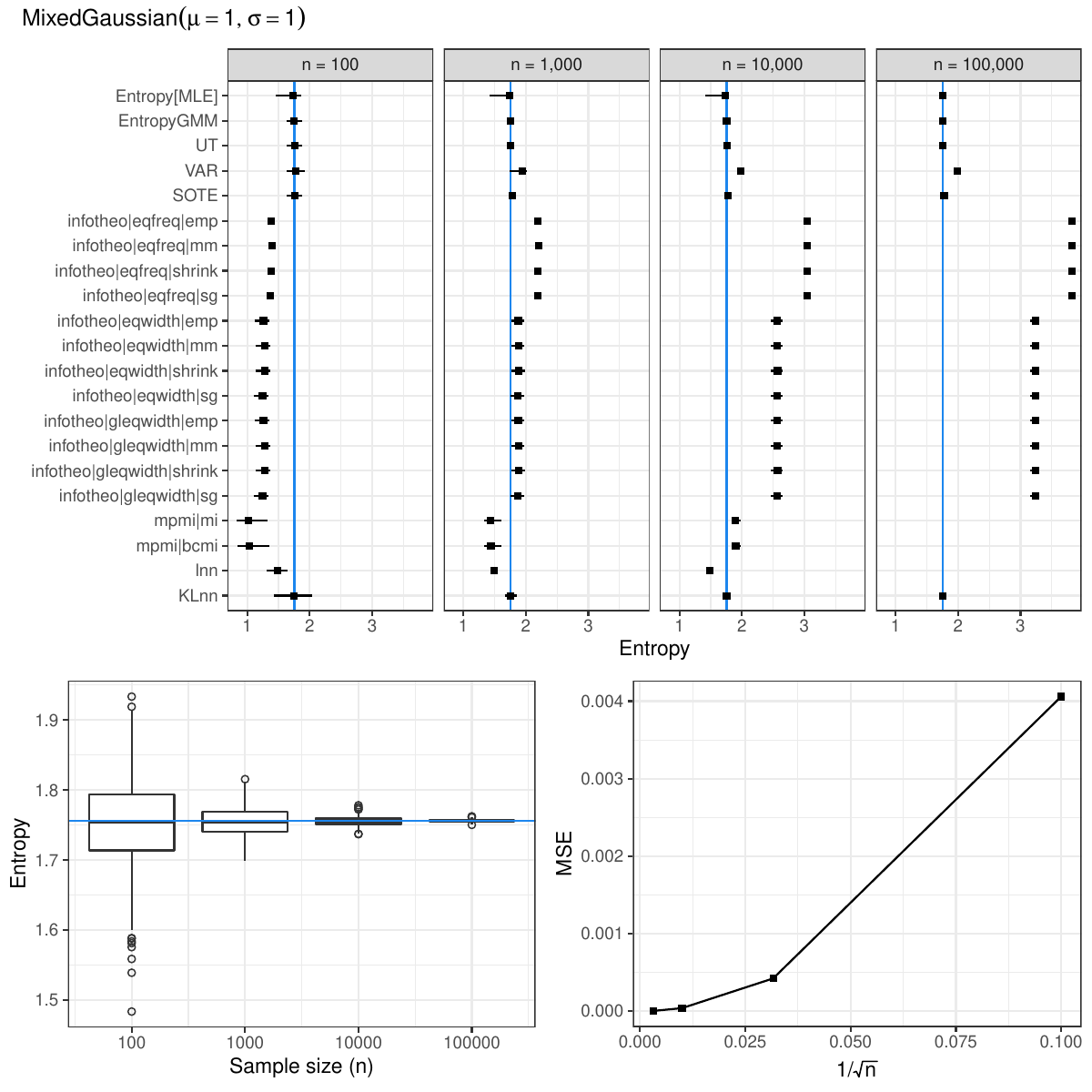}
\caption{\small Simulation results for univariate data generated from MixedGaussian($\mu= 1, \sigma = 1)$. For a description of each graph see Figure~\ref{fig2:mixedGauss}.}
\label{fig3:mixedGauss}
\end{figure}

\begin{figure}[htb]
\centering
\includegraphics[width=\textwidth]{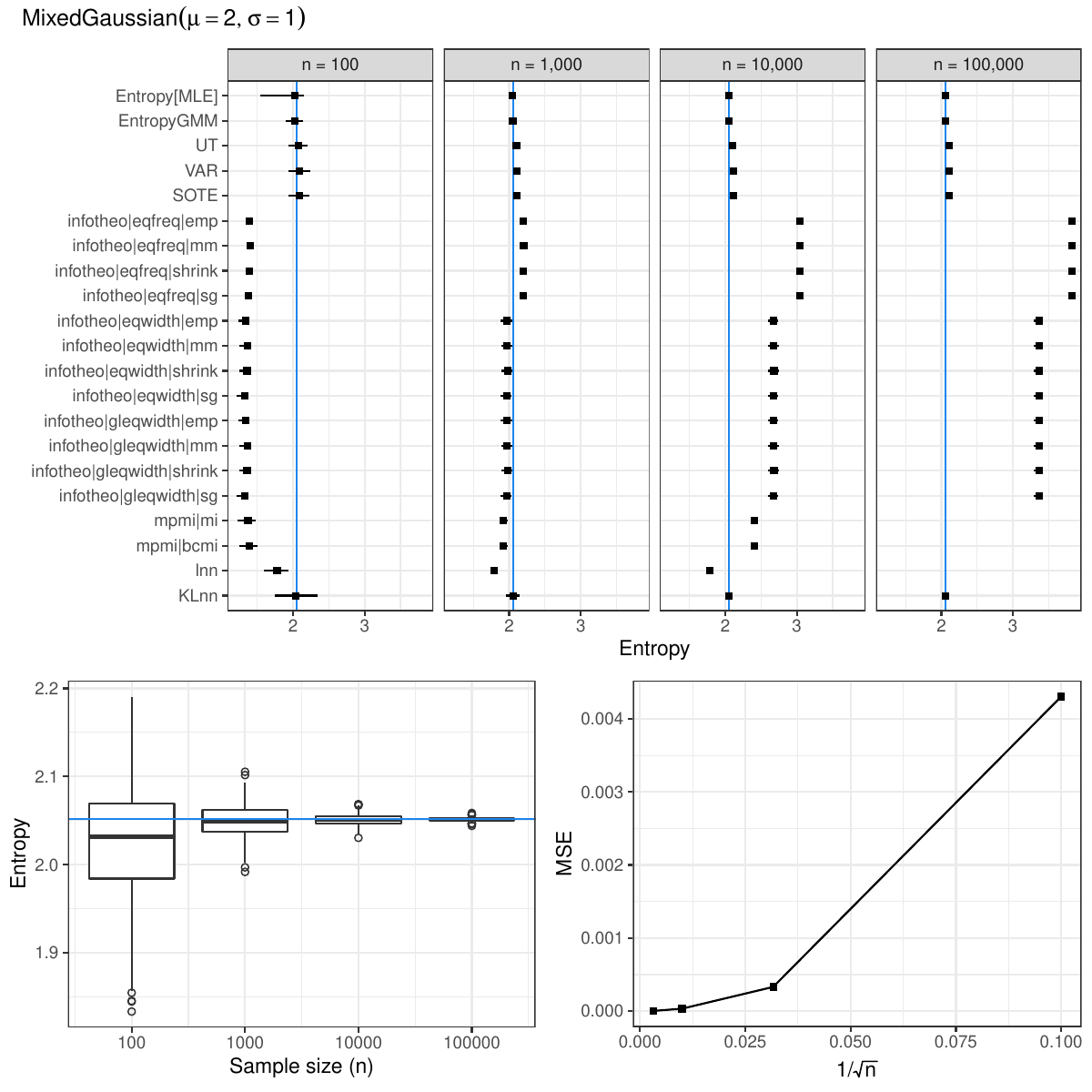}
\caption{\small Simulation results for univariate data generated from MixedGaussian($\mu= 2, \sigma = 1)$. For a description of each graph see Figure~\ref{fig2:mixedGauss}.}
\label{fig4:mixedGauss}
\end{figure}

In the simulation experiment we generated data from a set of mixed-Gaussian distributions with mean $\mu= \{0,1,2,3\}$ and $\sigma=1$. 
Figures~\ref{fig2:mixedGauss}--\ref{fig4:mixedGauss} show the results based on 1000 replications for increasing sample sizes.
Our estimator appears slightly biased for the smaller sample size, but it becomes quickly unbiased as $n$ grows and with decreasing variance, so there is a clear evidence of asymptotic efficiency (see the graphs at the bottom of  Figures~\ref{fig2:mixedGauss}--\ref{fig4:mixedGauss}). 
On the contrary, most nonparametric estimators appear to be highly biased, often underestimating for small sample sizes and overestimating for higher sample sizes, with the exception of the local nearest neighbour-based estimator ({\sf lnn}) which is always underestimating the true value of the entropy. Finally, note that for the larger sample size both the kernel-based and the nearest neighbour-based estimators could not be computed.

\begin{figure}[htb]
\centering
\includegraphics[width=\textwidth]{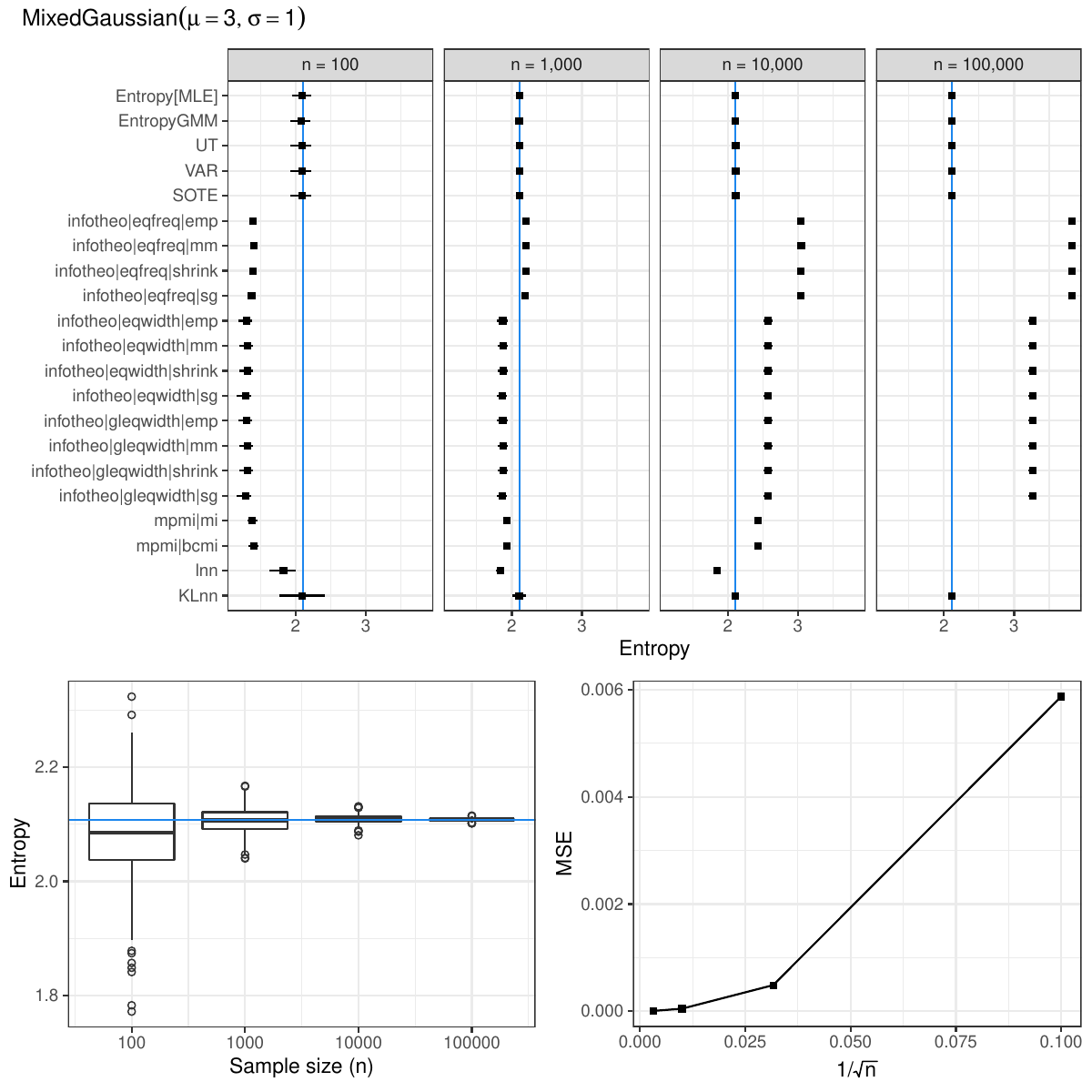}
\caption{\small Simulation results for univariate data generated from MixedGaussian($\mu= 3, \sigma = 1)$. For a description of each graph see Figure~\ref{fig2:mixedGauss}.}
\label{fig5:mixedGauss}
\end{figure}

\clearpage

\subsection{Laplace distribution}
\label{sec:laplace:dist}

\begin{figure}[htb]
\centering
\includegraphics[width=\textwidth]{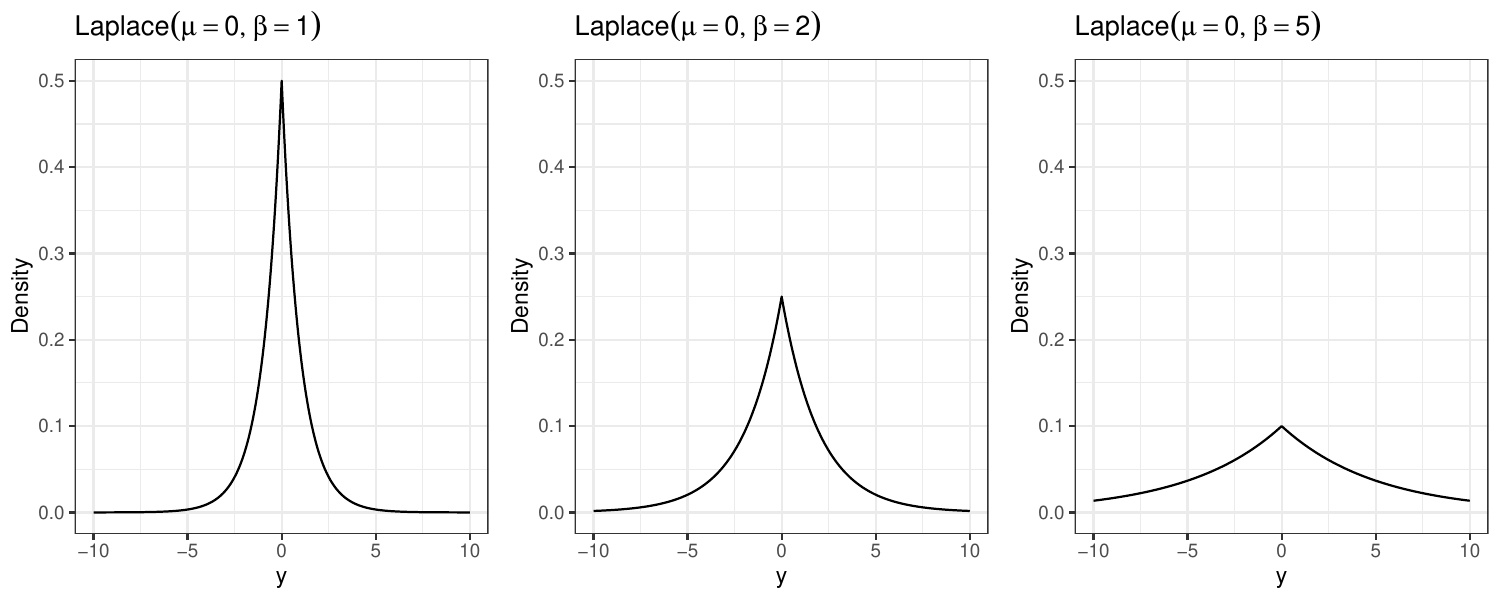}
\caption{\small Density plots of selected Laplace distributions.}
\label{fig1:laplace}
\end{figure}

In this second simulation experiment we generated data from the Laplace distribution, a symmetric but otherwise peaked and with fatter tails compared to the Gaussian distribution. Three parameter settings are considered as shown in Figure~\ref{fig1:laplace}.

\begin{figure}[htb]
\centering
\includegraphics[width=\textwidth]{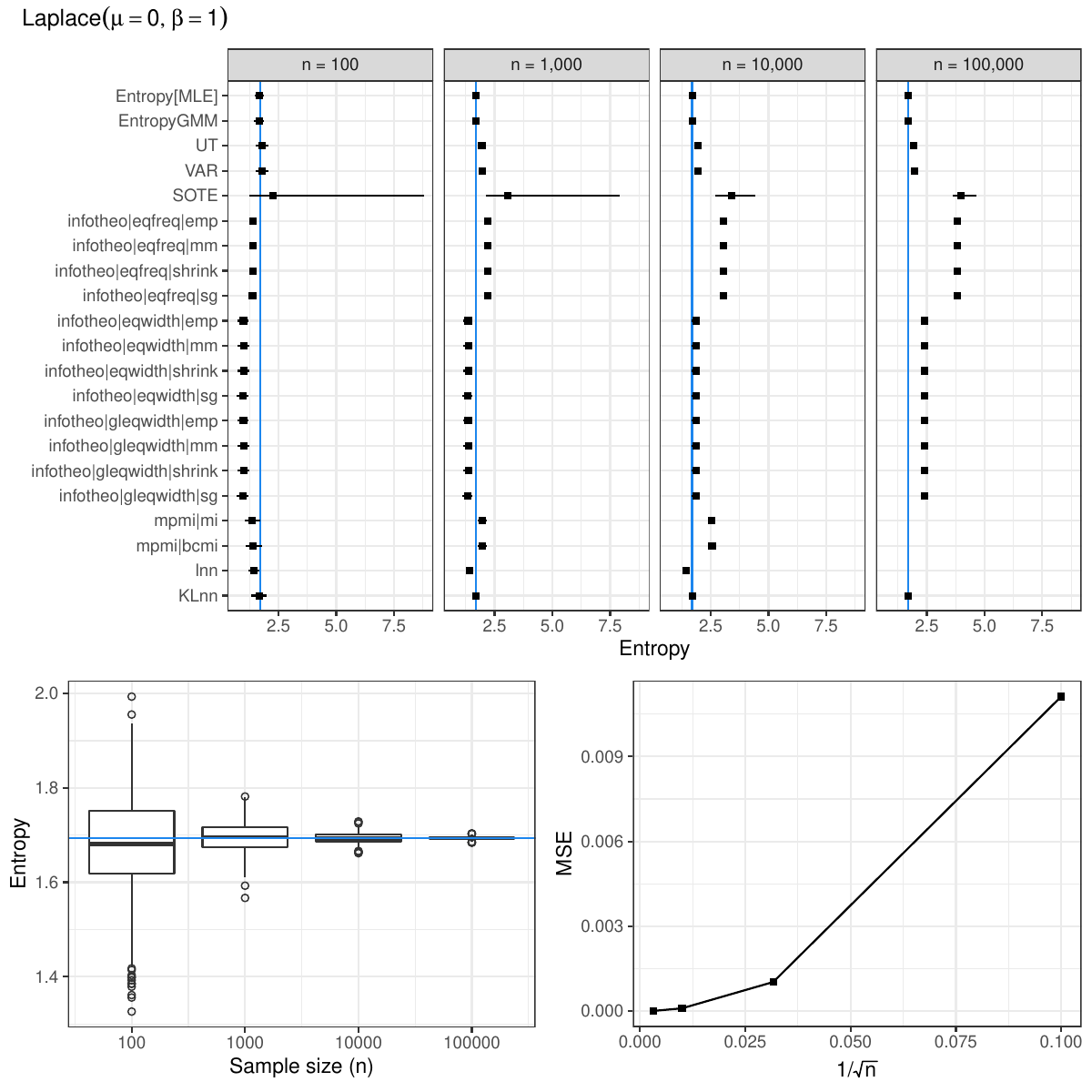}
\caption{\small Simulation results for univariate data generated from Laplace($\mu = 0, \beta = 1)$. For a description of each graph see Figure~\ref{fig2:mixedGauss}.}
\label{fig2:laplace}
\end{figure}

\begin{figure}[htb]
\centering
\includegraphics[width=\textwidth]{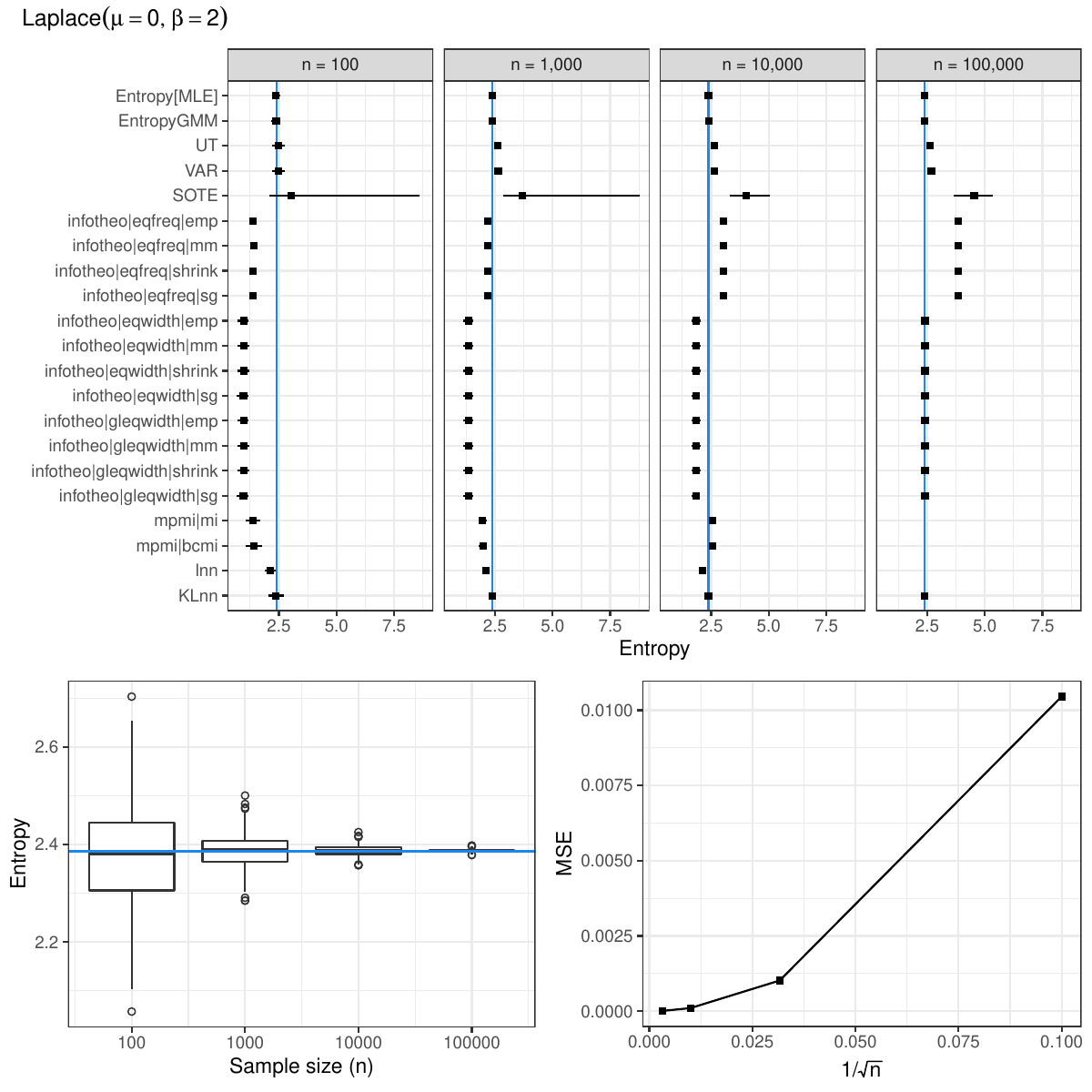}
\caption{\small Simulation results for univariate data generated from Laplace($\mu = 0, \beta = 2)$. For a description of each graph see Figure~\ref{fig2:mixedGauss}.}
\label{fig3:laplace}
\end{figure}

\begin{figure}[htb]
\centering
\includegraphics[width=\textwidth]{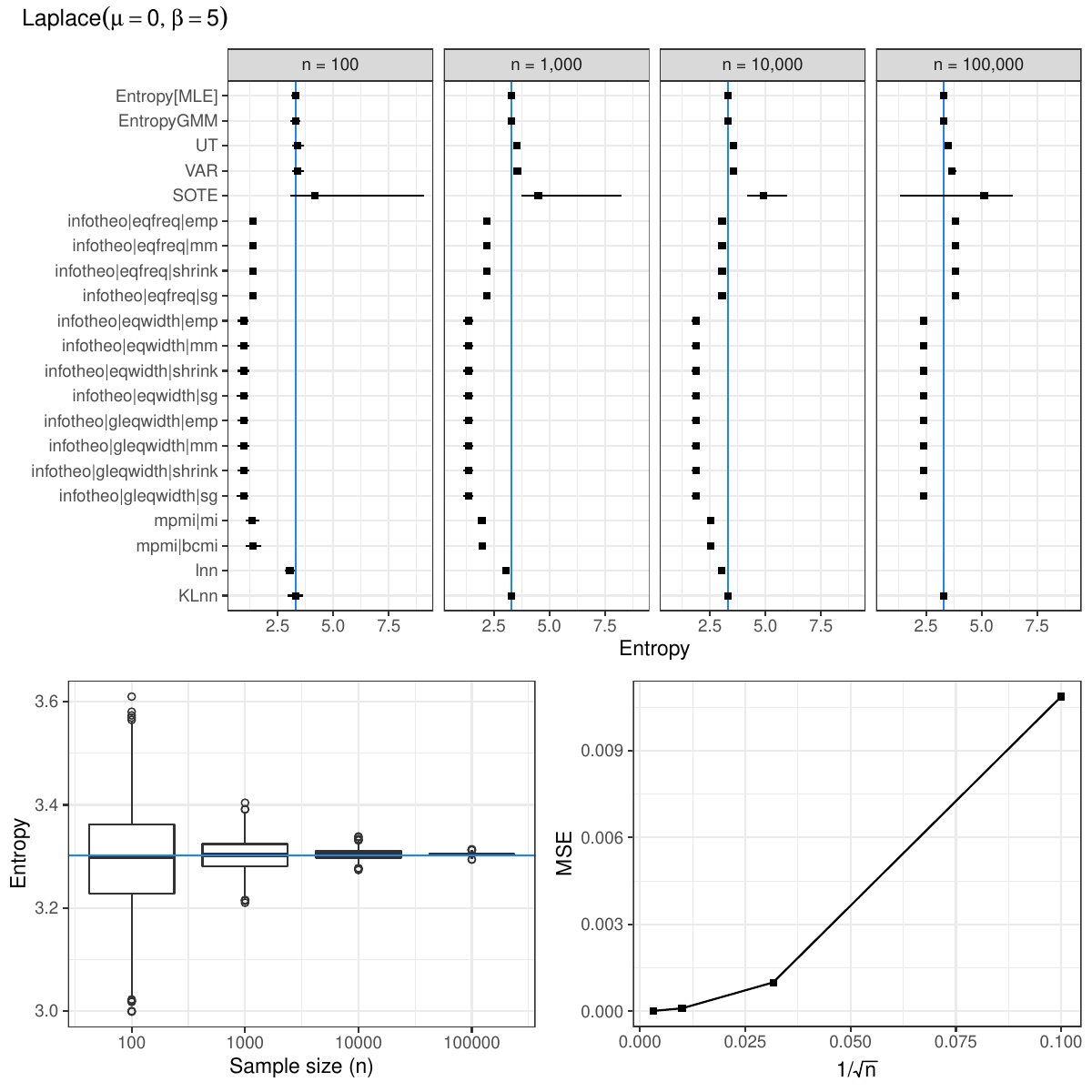}
\caption{\small Simulation results for univariate data generated from Laplace($\mu = 0, \beta = 5)$. For a description of each graph see Figure~\ref{fig2:mixedGauss}.}
\label{fig4:laplace}
\end{figure}

As in previous section, results based on 1000 replications for increasing sample sizes are shown in Figures~\ref{fig2:laplace}--\ref{fig4:laplace}. 
Again our estimator appears slightly biased for the smaller sample size, but it appears to be unbiased and efficient asymptotically. 
All nonparametric estimators are highly biased, except for the KLnn estimator, which however seems to have a larger variance in the smaller sample size case. 

\clearpage

\subsection{Bivariate Gaussian distribution} \label{sec:gauss:dist}

Assume a single-component multivariate Gaussian mixture, i.e. $f(y) = \Normal(y; \mub, \Sigmab)$. In this simple case, the MLE of the entropy is obtained by substituting the usual MLE estimates of $\mub$ and $\Sigmab$ in equation~\eqref{eq:entGauss}:
$$
\hat{\Entropy}(Y) = \frac{1}{2} \log \left( (2\pi e)^p |\hat{\Sigmab}| \right).
$$

\begin{figure}[htb]
\centering
\includegraphics[width=\textwidth]{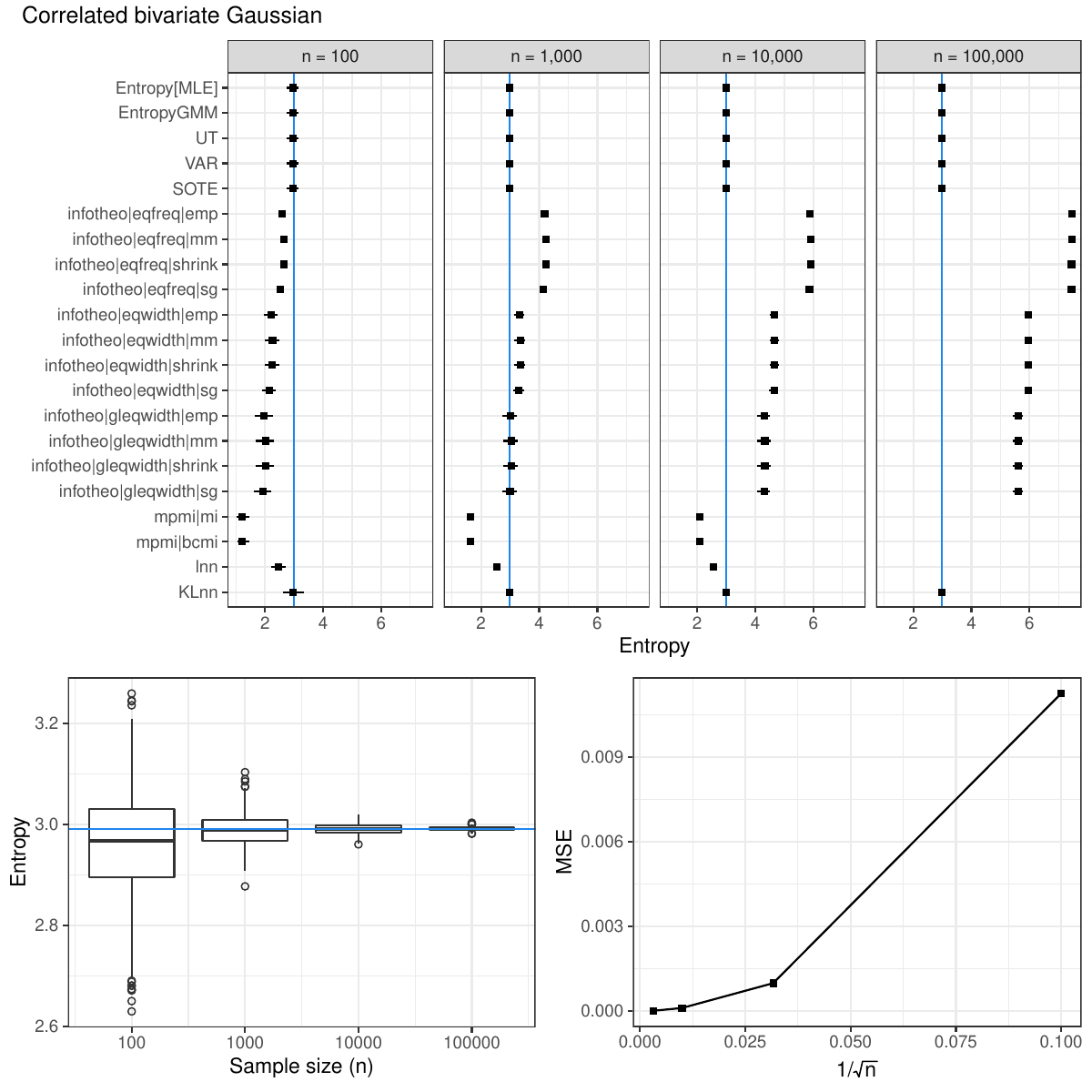}
\caption{\small Simulation results for correlated bivariate Gaussian distribution. For a description of each graph see Figure~\ref{fig2:mixedGauss}}
\label{fig:bivgauss}
\end{figure}

To investigate the sampling behaviour of our estimator, we conducted a simulation experiment where we generated data from a bivariate Gaussian distribution with mean $\mub = (0,0)\T$ 
and covariance matrix $\Sigmab = \begin{bsmallmatrix} 1.0 & 0.8 \\ 0.8 & 2.0 \end{bsmallmatrix}$. For each simulated dataset we compare the true entropy of the multivariate Gaussian with the estimates obtained by (i) plug-in the MLE of the covariance matrix, and (ii) using equation \eqref{eq:est} with estimates of unknown quantities obtained from the ``best'' GMM as selected by BIC. 
The box-plots in Figure~\ref{fig:bivgauss} show the results for 1000 replications. Our mixture-based estimator for the entropy is producing estimates almost identical to the MLEs obtained under the assumption of knowing the true data generating mechanism. There is a slight difference for sample size $n=100$, where the mixture-based estimator incorporates also the uncertainty due to the model selection step. However, the proposed estimator appears to be asymptotically unbiased and efficient. 

\clearpage

\subsection{Multivariate independent Chi-squared distribution}
\label{sec:chisq:dist}

Assume a 10-dimensional independent $\chi^2$ distribution with 5 degrees of freedom for each. dimension. 
Recall that for a $d$-dimensional independent random variable the entropy of the joint distribution is equal to the sum of the marginal entropies, i.e.
$$
\Entropy(Y) = \sum_{j=1}^d \Entropy(Y_j).
$$
If $Y_j \sim \chi^2(\nu)$, then the entropy is equal to 
$$
\Entropy(Y_j) = \log(2) + \log\Gamma(\nu/2) + \nu/2 + (1-\nu/2) \psi(\nu/2),
$$
where $\Gamma(x)$ is the gamma function, and $\psi(x) = \d \log\Gamma(x)/\d x$ the logarithmic derivative of the gamma function.
Thus, in our simulation setting the theoretical value of the entropy is equal to $\Entropy(Y) = 10 \times 2.423095 = 24.23095$.

\begin{figure}[htb]
\centering
\includegraphics[width=\textwidth]{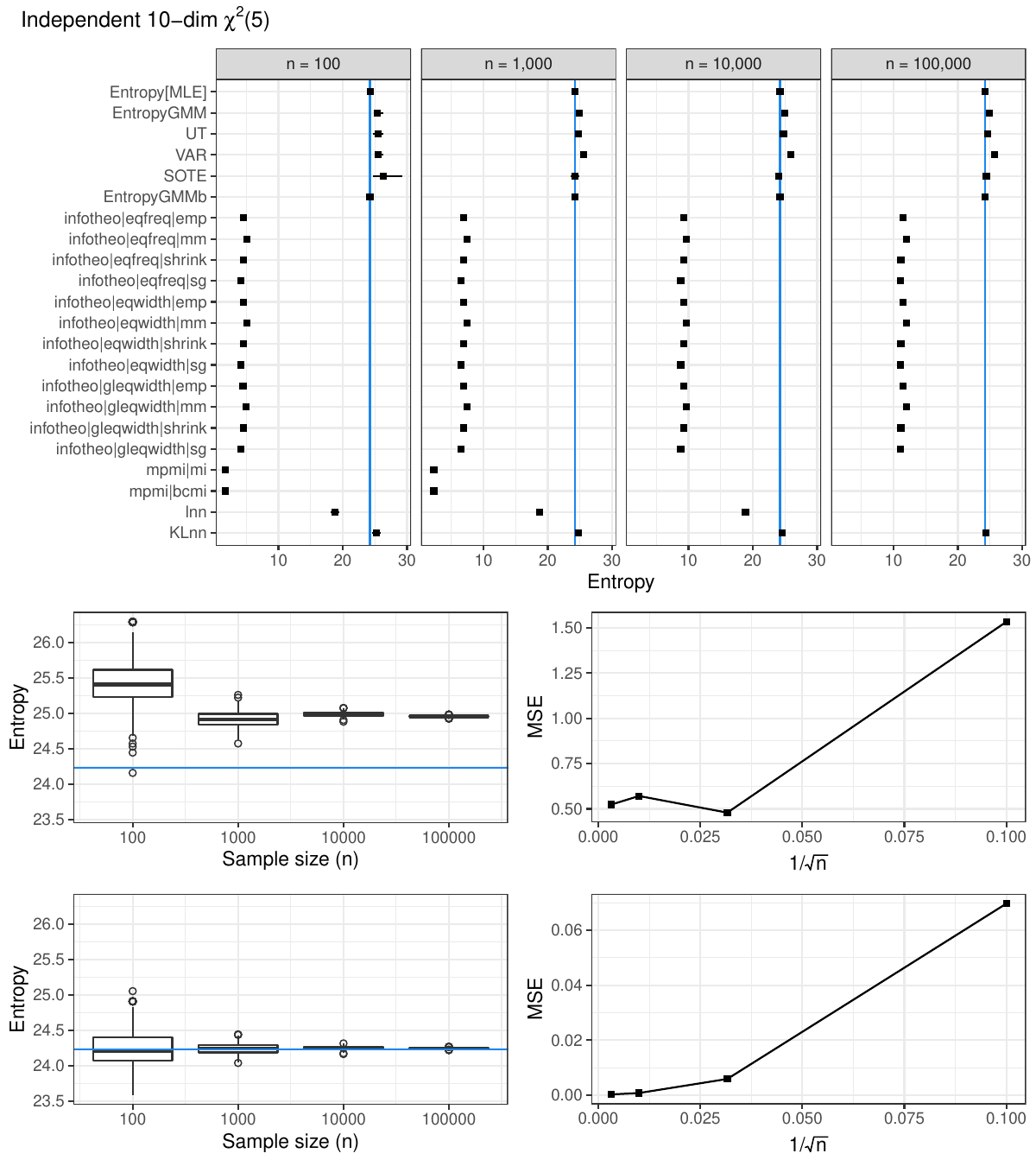}
\caption{\small Simulation results for the 10-dimensional independent Chi-squared distribution. For a description of each graph see Figure~\ref{fig2:mixedGauss}. The pair of graphs in the center refers	to the standard GMM-based estimator, whereas the pair of plots at the bottom refers to the GMM-based estimator using the density estimate for bounded data.}
\label{fig:ChiSquared}
\end{figure}

Figure~\ref{fig:ChiSquared} summarizes the results of the simulation experiments. In the multivariate case all the histogram-based and kernel-based estimators appear to be highly biased, and the same also happens, although to a lesser extent, for the nearest neighbour-based estimator. Compared to these, the entropy estimators based on GMMs are much closer on average to the true value, but they show a slight degrees of bias which does not vanish as the sample size increases. This seems to be related to the fact that we are approximating a multivariate distribution whose support is bounded at zero with an unbounded density arising from GMMs. To support this intuition we apply the proposed procedure for entropy estimation using the GMM-transformation approach for bounded data proposed by \cite{Scr19}. This is indicated as \EntropyGMMb\ and, as it can be clearly seen, the improvement coming from the density estimation step yields an unbiased and efficient estimator of the true entropy.

\clearpage

\subsection{A comparison in terms of efficiency and computing effort}
\label{sec:comparison}

The simulation studies reported in the previous sections show that the mixture-based estimate of the entropy is consistent and accurate, with only the Kozachenko \& Leonenko estimator providing a comparable accuracy. 
Table~\ref{tab:rmse} provides a direct comparison between the two methods in terms of the root mean square error (RMSE). Overall the Gaussian mixture-based estimator appears to be about 50\% more efficient than the Kozachenko \& Leonenko nearest neighbour estimator.
\SR{}{In the case of a bounded distribution, KLnn turns out to be more accurate than EntropyGMM, but much less that its bounded version EntropyGMMb.}
However, we should note that it is computationally more intensive, in particular if a mixture model needs to be estimated. 
Table~\ref{tab:systime} reports the system time (in seconds) required by the two estimators, and by the mixture-based estimator assuming that the Gaussian mixture model has already been estimated.

\begin{table}[ht]
\centering
\caption{\small RMSE for the GMM-based estimator of the entropy (\EntropyGMM) and the Kozachenko \& Leonenko nearest neighbour estimator (\KLnn). In the last case, the results reported also include to the GMM-based estimator for bounded data (\EntropyGMMb).}
\label{tab:rmse}
\setlength{\tabcolsep}{10pt}
\begin{tabular}{lrrrr}
\hline
Sample size &    100 &   1,000 &  10,000 & 100,000 \\
\hline
& \multicolumn{4}{c}{MixedGaussian($\mu=0,\sigma=1$)} \\
\cline{2-5}
\EntropyGMM & 0.0732 & 0.0225 & 0.0072 & 0.0023 \\
\KLnn       & 0.1665 & 0.0525 & 0.0167 & 0.0052 \\
\hline
& \multicolumn{4}{c}{MixedGaussian($\mu=1,\sigma=1$)} \\
\cline{2-5}
\EntropyGMM & 0.0637 & 0.0206 & 0.0063 & 0.0018 \\
\KLnn       & 0.1583 & 0.0492 & 0.0158 & 0.0049 \\
\hline
& \multicolumn{4}{c}{MixedGaussian($\mu=2,\sigma=1$)} \\
\cline{2-5}
\EntropyGMM & 0.0656 & 0.0182 & 0.0059 & 0.0019 \\
\KLnn       & 0.1581 & 0.0483 & 0.0156 & 0.0051 \\
\hline
& \multicolumn{4}{c}{MixedGaussian($\mu=3,\sigma=1$)} \\
\cline{2-5}
\EntropyGMM & 0.0766 & 0.0221 & 0.0069 & 0.0022 \\
\KLnn       & 0.1670 & 0.0518 & 0.0158 & 0.0051 \\
\hline
& \multicolumn{4}{c}{Laplace($\mu=0,\beta=1$)} \\
\cline{2-5}
\EntropyGMM & 0.1054 & 0.0321 & 0.0102 & 0.0033 \\
\KLnn       & 0.1728 & 0.0543 & 0.0175 & 0.0057 \\
\hline
& \multicolumn{4}{c}{Laplace($\mu=0,\beta=2$)} \\
\cline{2-5}
\EntropyGMM & 0.1022 & 0.0320 & 0.0103 & 0.0033 \\
\KLnn       & 0.1774 & 0.0556 & 0.0175 & 0.0054 \\
\hline
& \multicolumn{4}{c}{Laplace($\mu=0,\beta=5$)} \\
\cline{2-5}
\EntropyGMM & 0.1042 & 0.0316 & 0.0101 & 0.0034 \\
\KLnn       & 0.1757 & 0.0548 & 0.0173 & 0.0055 \\
\hline
& \multicolumn{4}{c}{Correlated bivariate Gaussian} \\
\cline{2-5}
\EntropyGMM & 0.1061 & 0.0315 & 0.0105 & 0.0033 \\
\KLnn       & 0.1831 & 0.0567 & 0.0185 & 0.0059 \\
\hline
& \multicolumn{4}{c}{Independent Chi-squared} \\
\cline{2-5}
\EntropyGMM  & 1.2383 & 0.6919 & 0.7553 &  0.7233 \\
\EntropyGMMb & 0.2640 & 0.0768 & 0.0281 &  0.0165 \\
\KLnn        & 1.0265 & 0.5290 & 0.3242 &  0.1985 \\
\hline
\end{tabular}
\end{table}

\begin{table}[ht]
\centering
\caption{\small System time (in seconds) required by the GMM-based estimator of the entropy (\EntropyGMM), the Kozachenko \& Leonenko nearest neighbour estimator (\KLnn), and by the GMM-based estimator assuming that the mixture model is already available (\EntropyGMM*). Moreover, in the last case the results reported refer to the GMM-based estimator for bounded data (\EntropyGMMb).}
\label{tab:systime}
\setlength{\tabcolsep}{10pt}
\begin{tabular}{lrrrr}
\hline
Sample size &    100 &   1,000 &  10,000 & 100,000 \\
\hline
& \multicolumn{4}{c}{MixedGaussian($\mu=0,\sigma=1$)} \\
\cline{2-5}
\EntropyGMM  & 0.0260 & 0.2061 & 1.0190 & 10.1114 \\
\EntropyGMM* & 0.0001 & 0.0001 & 0.0003 &  0.0020 \\
\KLnn        & 0.0002 & 0.0010 & 0.0125 &  0.1408 \\
\hline
& \multicolumn{4}{c}{MixedGaussian($\mu=1,\sigma=1$)} \\
\cline{2-5}
\EntropyGMM  & 0.0214 & 0.0828 & 0.8869 & 8.4041 \\
\EntropyGMM* & 0.0001 & 0.0001 & 0.0003 & 0.0024 \\
\KLnn        & 0.0002 & 0.0010 & 0.0099 & 0.1310 \\
\hline
& \multicolumn{4}{c}{MixedGaussian($\mu=2,\sigma=1$)} \\
\cline{2-5}
\EntropyGMM  & 0.0255 & 0.0769 & 0.7414 & 7.3904 \\
\EntropyGMM* & 0.0001 & 0.0001 & 0.0004 & 0.0018 \\
\KLnn        & 0.0002 & 0.0009 & 0.0094 & 0.1347 \\
\hline
& \multicolumn{4}{c}{MixedGaussian($\mu=3,\sigma=1$)} \\
\cline{2-5}
\EntropyGMM  & 0.0160 & 0.0942 & 0.7337 & 6.9345 \\
\EntropyGMM* & 0.0001 & 0.0001 & 0.0004 & 0.0023 \\
\KLnn        & 0.0002 & 0.0008 & 0.0093 & 0.1324 \\
\hline
& \multicolumn{4}{c}{Laplace($\mu=0,\beta=1$)} \\
\cline{2-5}
\EntropyGMM  & 0.0241 & 0.1285 & 1.4072 & 13.7191 \\
\EntropyGMM* & 0.0001 & 0.0001 & 0.0007 &  0.0052 \\
\KLnn        & 0.0002 & 0.0009 & 0.0104 &  0.1353 \\
\hline
& \multicolumn{4}{c}{Laplace($\mu=0,\beta=2$)} \\
\cline{2-5}
\EntropyGMM  & 0.0151 & 0.1332 & 1.2136 & 12.5126 \\
\EntropyGMM* & 0.0001 & 0.0001 & 0.0005 &  0.0058 \\
\KLnn        & 0.0002 & 0.0009 & 0.0089 &  0.1323 \\
\hline
& \multicolumn{4}{c}{Laplace($\mu=0,\beta=5$)} \\
\cline{2-5}
\EntropyGMM  & 0.0157 & 0.1117 & 0.9766 & 11.8614 \\
\EntropyGMM* & 0.0001 & 0.0001 & 0.0007 &  0.0068 \\
\KLnn        & 0.0002 & 0.0009 & 0.0094 &  0.1790 \\
\hline
& \multicolumn{4}{c}{Correlated bivariate Gaussian} \\
\cline{2-5}
\EntropyGMM  & 0.2304 & 3.0034 & 16.7284 & 137.0333 \\
\EntropyGMM* & 0.0001 & 0.0001 &  0.0003 &   0.0020 \\
\KLnn        & 0.0003 & 0.0012 &  0.0131 &   0.1660 \\
\hline
& \multicolumn{4}{c}{Independent Chi-squared} \\
\cline{2-5}
\EntropyGMMb  & 18.9631 & 23.8822 & 65.7678 & 678.0250 \\
\EntropyGMMb* &  0.0000 &  0.0000 &  0.0001 &   0.0009 \\
\KLnn         &  0.0007 &  0.0147 &  0.5512 &  29.9988 \\
\hline
\end{tabular}
\end{table}

\clearpage


\subsection{Estimating mutual information from multivariate log-normal distribution}
\label{sec:mi}

Mutual information (MI) is another important concept in information theory that is especially useful when trying to decipher the dependency structure that exists between a set of variables. The mutual information between two variables $Y_1$ and $Y_2$ relates to their marginal and joint entropies as
$
MI(Y_1, Y_2) = \Entropy(Y_1) + \Entropy(Y_2) - \Entropy(Y_1, Y_2).
$
A straightforward plugging-in estimate is $\hat{MI}(Y_1, Y_2) = \hat{\Entropy}(Y_1) + \hat{\Entropy}(Y_2) - \hat{\Entropy}(Y_1, Y_2)$.

The multivariate log-normal distribution is among the few multivariate distributions for which a closed-form expression of the entropy is available. Indeed, if $Y$ is a $p$-dimensional random log-normal vector with parameter $(\mub, \Sigmab)$, \cite{QSL16} showed that
$$
\Entropy(Y) = \frac{p}{2} (1 + \log 2 \pi) + \frac{1}{2} \log |\Sigmab| + \sum_{j=1}^p \mu_j.
$$
To use the multivariate log-normal as a benchmark {for assessing} the accuracy of our estimate, we need to account for the fact that its support is restricted to $\Real_+^p$.
To this aim, we resort to the adaptation of GMM to bounded distributions proposed by \cite{Scr19}.

To assess the accuracy of the GMM and bounded GMM estimates of the mutual information, we sampled $B=500$ bivariate random variables with mean $(\mu_1, \mu_2)$, variances $(\sigma^2_1, \sigma^2_2)$ and correlation $\rho$. We considered two configurations with low means, so that the boundedness clearly affects the distribution: $(\mu_1 = \mu_2 = 0, \sigma^2_1=1, \sigma^2_2=1/4)$ and $(\mu_1 = \mu_2 = 1, \sigma^2_1=1, \sigma^2_2=2)$. We consider three levels of correlation $\rho = 0.1, 0.5, \text{and } 0.9$ for both configurations, to control the mutual information to be estimated.

\begin{figure}[htb]
\centering
\includegraphics[width=\textwidth]{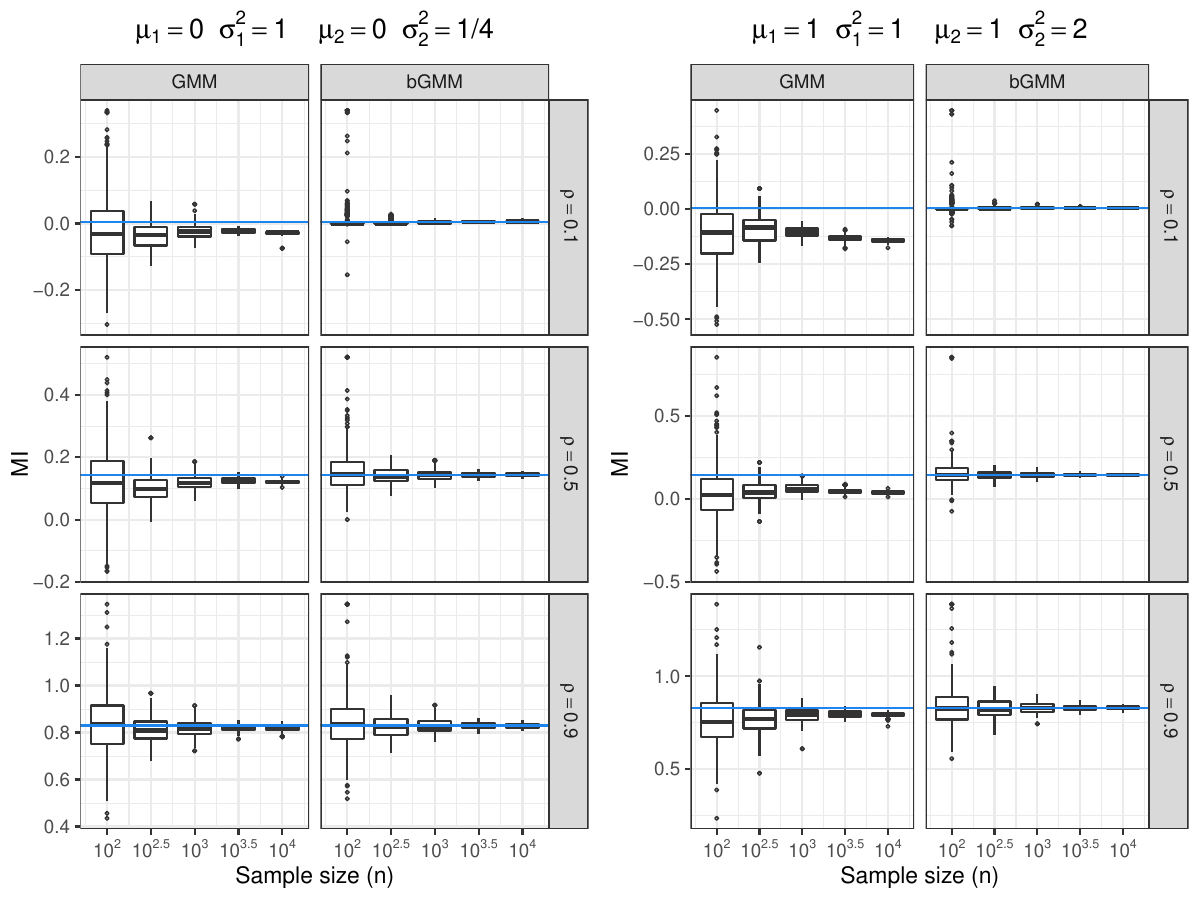}
\caption{\small Distribution of estimates of the mutual information of a bivariate log-normal variables with GMM and bounded GMM (bGMM) over 500 replicates for two configurations of means and variances, different correlation coefficients, and varying sample size. Horizontal blue lines refer to the true theoretical value.}
\label{fig:simulMI}
\end{figure}

Figure \ref{fig:simulMI} clearly shows that the use of a standard GMM induces a systematic bias in the estimation, which does not vanish when the sample size increases. It also shows that resorting to the bounded GMM corrects this bias, which is negligible, whatever the correlation level.

\clearpage

\section{Applications} \label{sec:app}
\subsection{Image quantization and segmentation} 
\label{sec:image}

\begin{figure}[htb]
\centering
\includegraphics[width=0.45\textwidth]{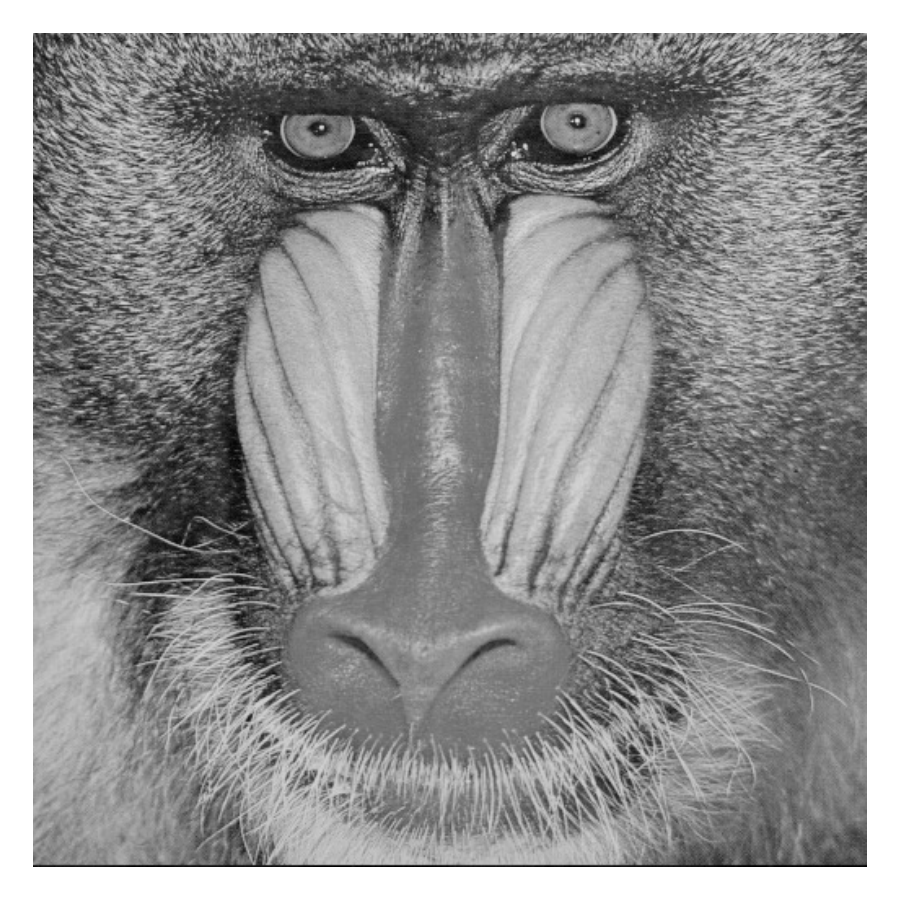}
\includegraphics[width=0.45\textwidth]{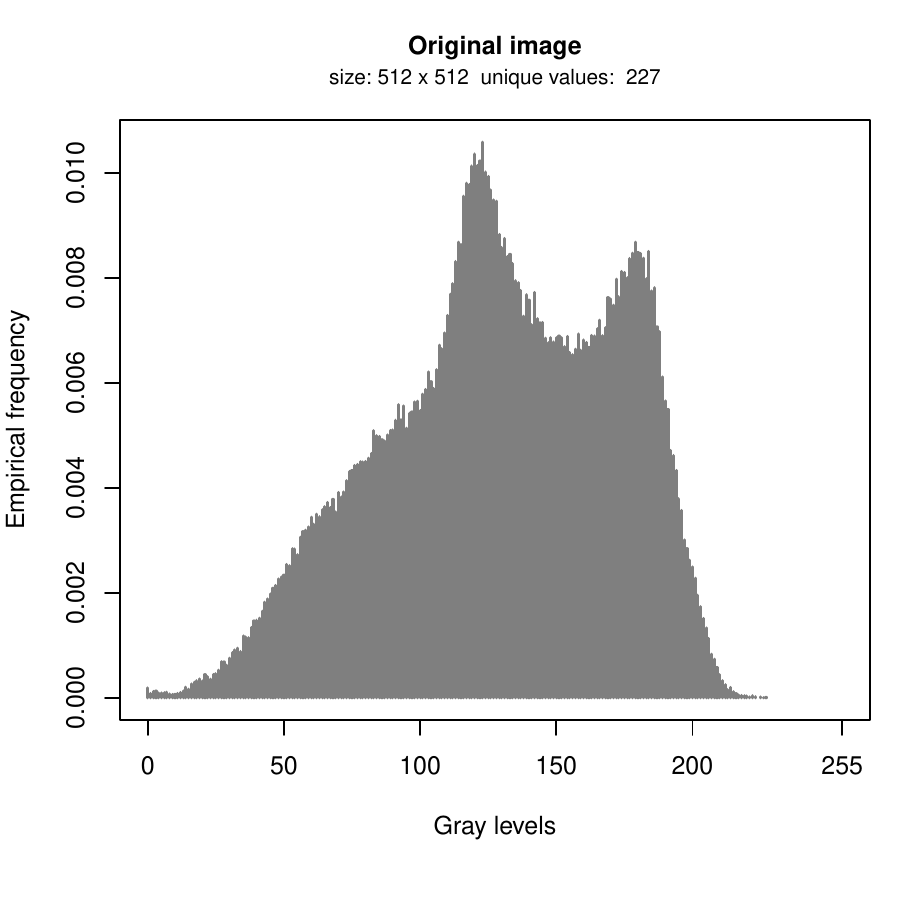}
\caption{\small Baboon black and white image (left panel) and corresponding histogram of grey levels (right panel).}
\label{fig:baboon}
\end{figure}

In image processing the term \emph{quantization} refers to the process of compressing a range of values to a single intensity value. The resulting image can then be partitioned into homogeneous areas, a process called \emph{image segmentation}. For instance, consider the black and white image of a baboon in the left panel of Figure~\ref{fig:baboon}. This is often used as benchmark in digital image processing and it is available on the USC-SIPI image database at \url{http://sipi.usc.edu/database}. The histogram in the right panel of Figure~\ref{fig:baboon} represents the distribution of grey level intensity values for each pixel of the image.

The main goal of digital image segmentation is to find a reduced version of the original image that captures the main characteristics. A density-based approach can be adopted by fitting a GMM with unconstrained variances to approximate the distribution of grey levels.
The entropy of an estimated GMM provides a measure of the uncertainty associated with the segmented image, which can then be compared with the entropy of the empirical distribution of intensity levels. 

\begin{figure}[htb]
\centering
\includegraphics[width=0.6\textwidth]{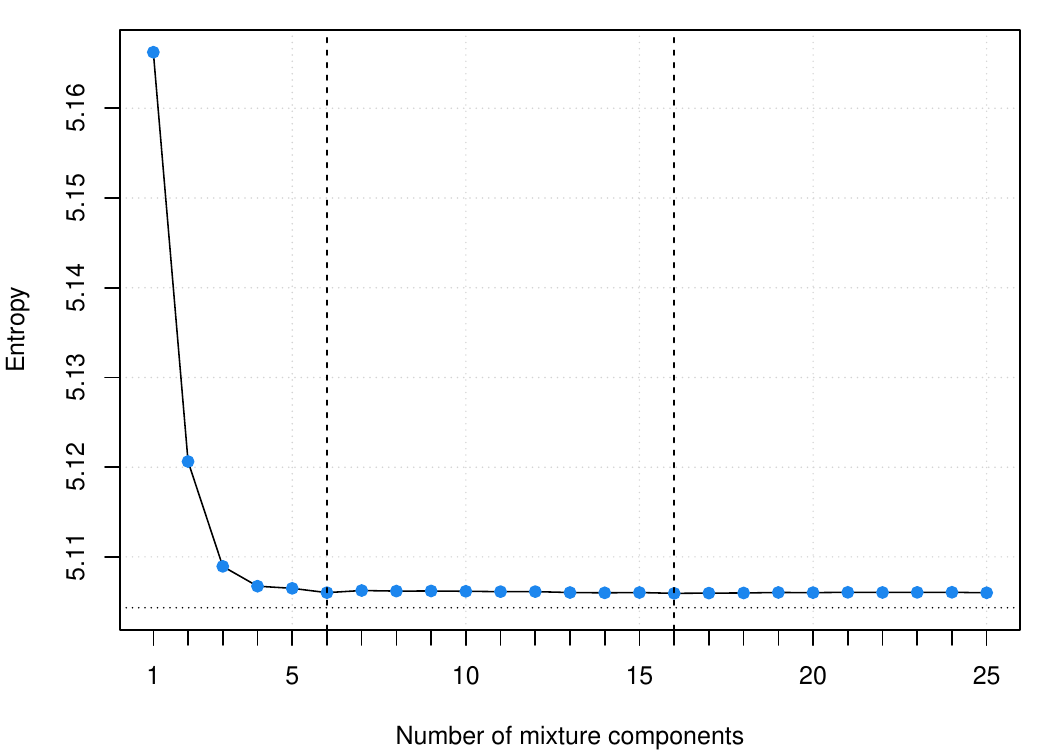}
\caption{\small Graph of entropy values for GMMs fitted to the grey levels of black and white baboon image with increasing number of mixture components. Vertical dashed lines represent the first local minimum and the global minimum of entropy, while the horizontal dotted line corresponds to the entropy of the empirical distribution of grey levels in the original image.}
\label{fig:baboon_entropy}
\end{figure}

Figure~\ref{fig:baboon_entropy} shows the values of the GMM entropy as a function of the number of mixture components.
As the number of components increases, the segmented image approaches the original image and, consequently, the entropy decreases.
Thus, the main objective becomes to select a segmented image that, simultaneously, uses a reduced number of mixture components and attains a small entropy value.

\begin{figure}[htb]
\centering
\includegraphics[width=0.45\textwidth]{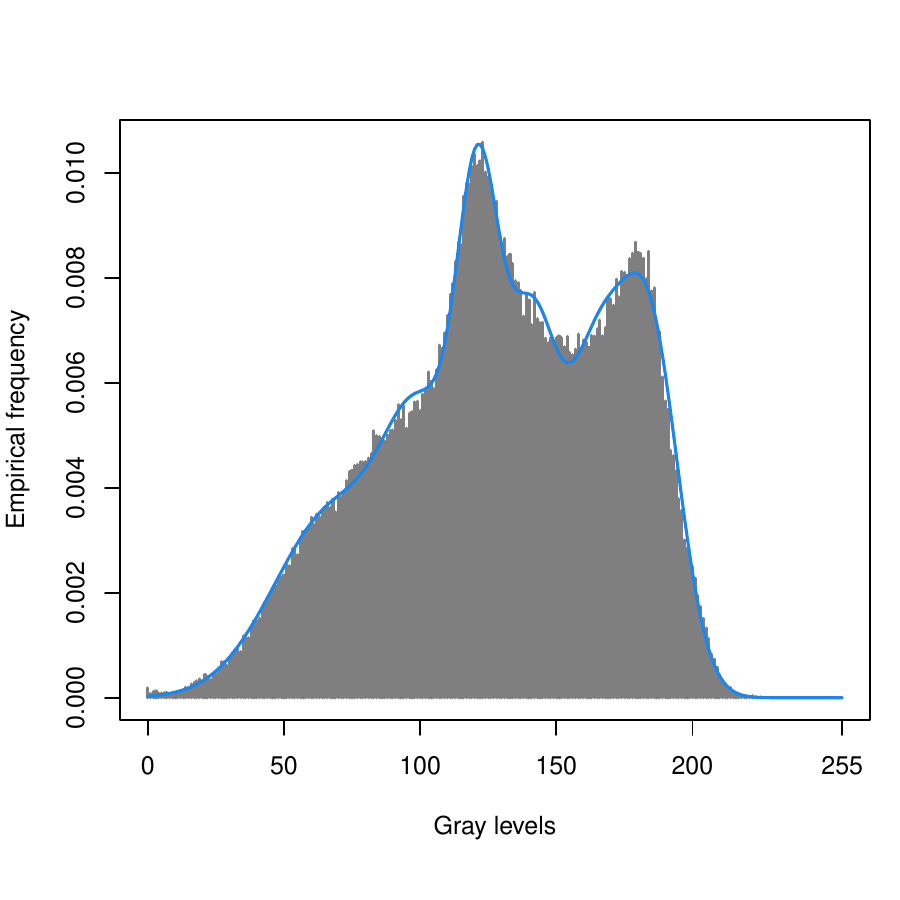}
\includegraphics[width=0.45\textwidth]{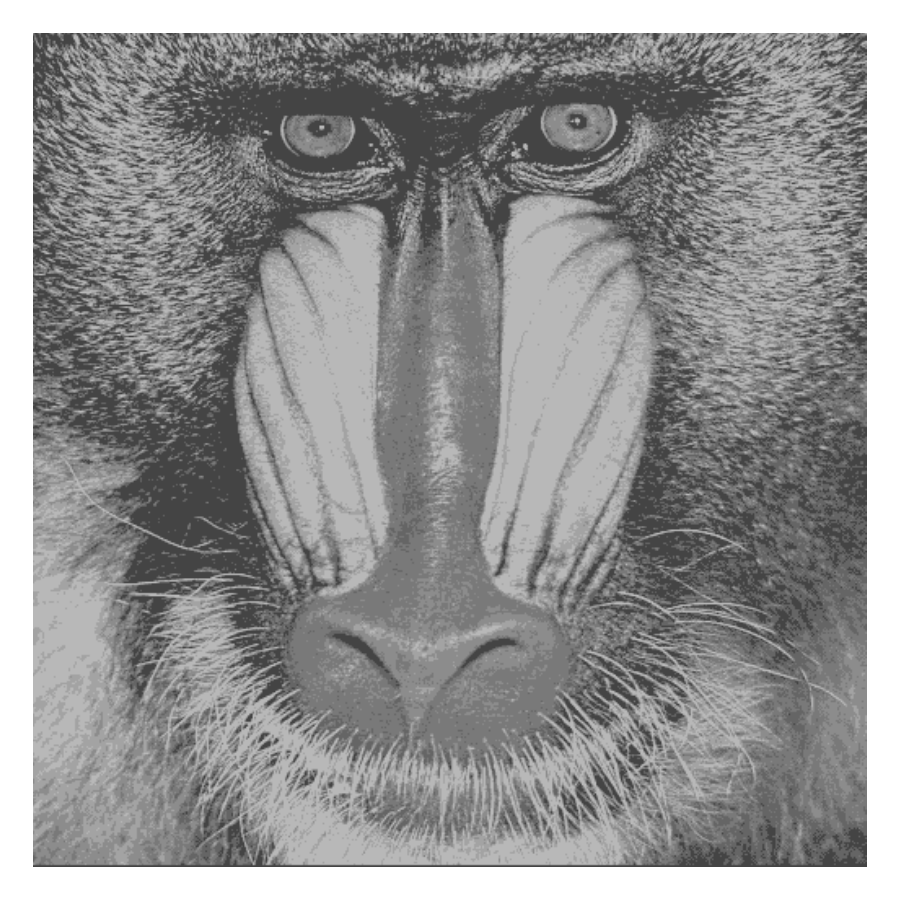}\\
\includegraphics[width=0.45\textwidth]{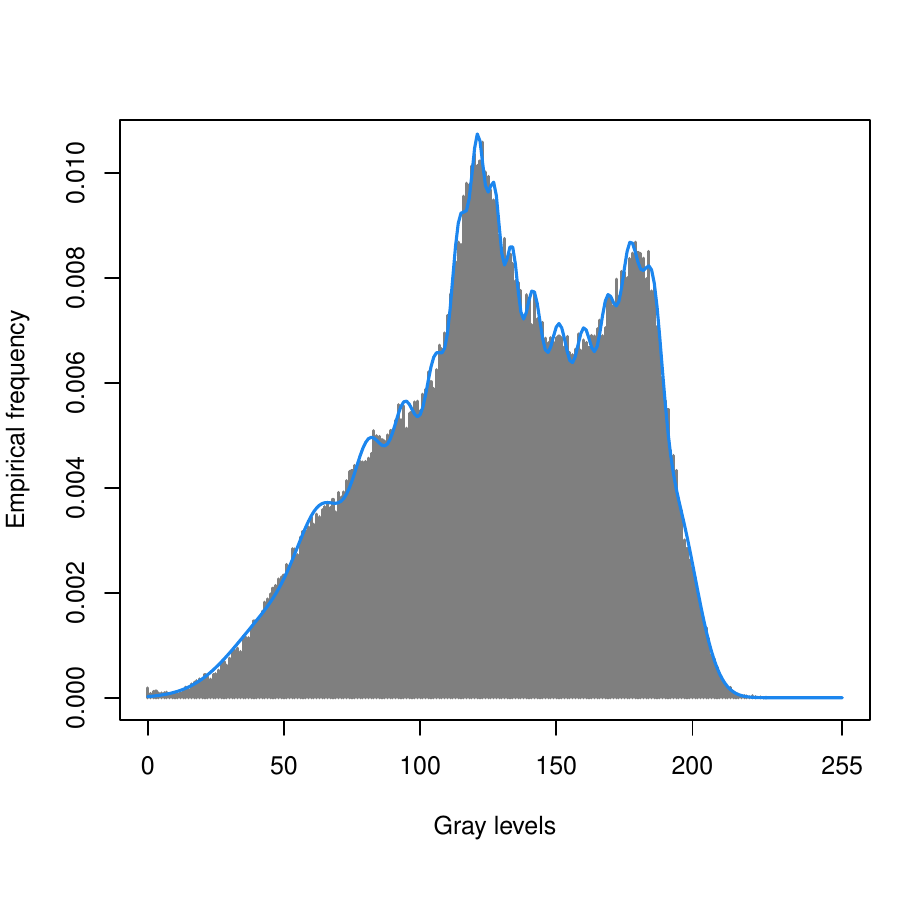}
\includegraphics[width=0.45\textwidth]{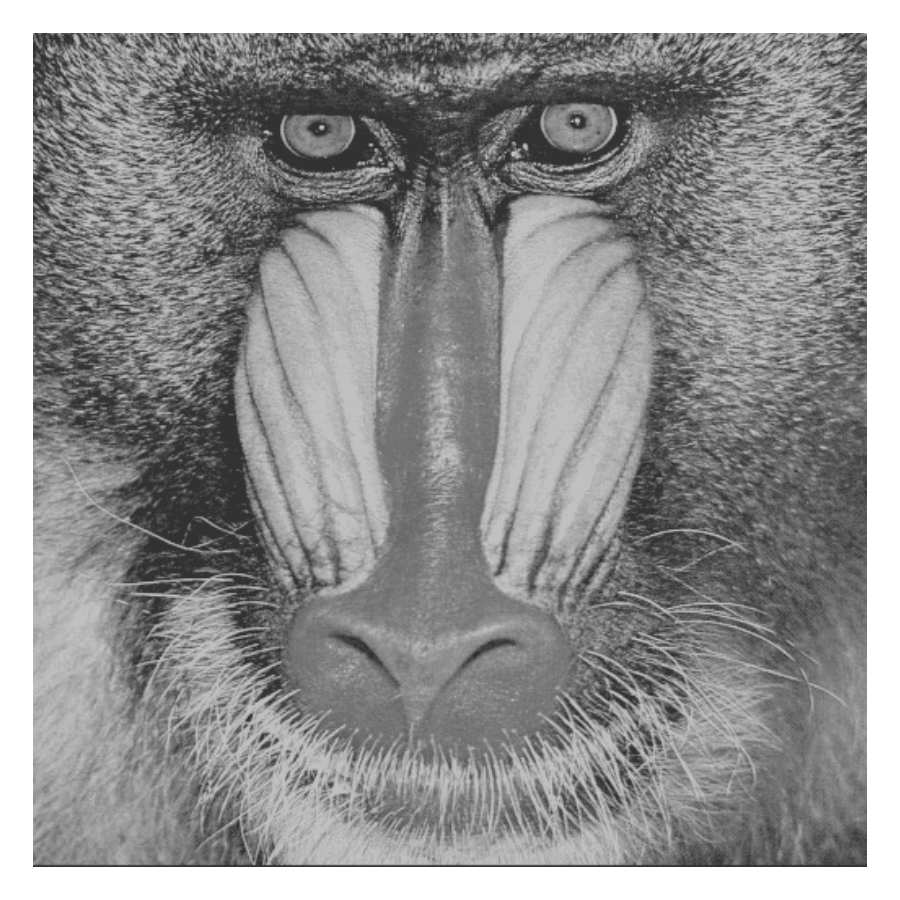}
\caption{\small Histograms with density estimates (left panels) obtained from GMMs with 6 (top) and 16 (bottom) mixture components, corresponding to the first local minimum and the global minimum, respectively, of the entropy. The images on the right panel show the associated segmented images.}
\label{fig:baboon_GMMs}
\end{figure}

We may consider the 6-component mixture model corresponding to the first local minimum, or the 16-component mixture model corresponding to the global minimum of the entropy. 
The histograms of intensity values with the estimated densities for the selected GMMs are shown in the left panels of Figure~\ref{fig:baboon_GMMs}, with the associated segmented images on the right. The solution with 6 mixture components is able to capture the main characteristics and shapes of the image. By selecting the 16-component GMM the segmented image is virtually indistinguishable from the original. 

These findings can be corroborated by computing the structural similarity (SSIM) index, which is a widely adopted index for measuring the similarity between two images and evaluating the perceived change in structural information \citep{Wang:etal:2004}. Table~\ref{tab:baboon_compression} reports the values of the SSIM index for the two solutions, showing that both segmented images have very high perceived quality.
Finally, we note that by applying GMMs for quantization we can achieve a significant image compression. In fact, the original image is an 8-bit image of dimension $512 \times 512$, so 2048 Kb are needed to store the image. The size of the segmented images shown in the right panels of Figure~\ref{fig:baboon_GMMs} and the corresponding compression rates are also reported in Table~\ref{tab:baboon_compression}.
Replacing the original image with the segmented image using 16 Gaussian components halves the size (CR = 2) with essentially the same visual perception (SSIM = 0.9939). By using the segmented image obtained with 6 Gaussian components the size is further reduced to about one third (CR = 3) with a slight loss in visual perception (SSIM = 0.9713).

\begin{table}[ht]
\centering
\caption{\small Comparison of original baboon image and segmented images obtained by fitting GMMs with 16 and 6 mixture components. Reported values are the entropy (smaller is better), the SSIM index (larger is better), the size and compression rate (CR) of images.}
\label{tab:baboon_compression}
\begin{tabular}{lccccc}
\hline
Image & Entropy & SSIM & Size (kb) & CR \\ 
\hline
Original         & 5.104336 & 1.0000 & 2048.00 & 1.00 \\ 
Segmented GMM-16 & 5.105943 & 0.9939 & 1024.00 & 2.00 \\ 
Segmented GMM-6  & 5.106022 & 0.9713 & \phantom{0}661.75 & 3.09 \\
\hline
\end{tabular}
\end{table}

\clearpage

\subsection{Inference of tree-shaped graphical models}
\label{sec:tree}


Graphical models \citep{Lau96} provide a generic framework to analyse the dependency structure within a set of variables. Graphical models are most often assumed to be sparse, meaning that most pairs of variables are supposed to be independent conditionally on all the others. Assuming that the graphical model is tree-shaped, \cite{ChL68} showed that the determination of the tree with maximum likelihood can be formulated as a maximum spanning tree (MST) problem, using the mutual information between each pair of variables as the edge weight.

\begin{figure}[htb]
  \centering
  \begin{tabular}{cc|ccc}
    \multicolumn{2}{c|}{Climate indices} & 
    \multicolumn{3}{c}{Telescope components} \\
    \hline
    \includegraphics[width=.15\textwidth]{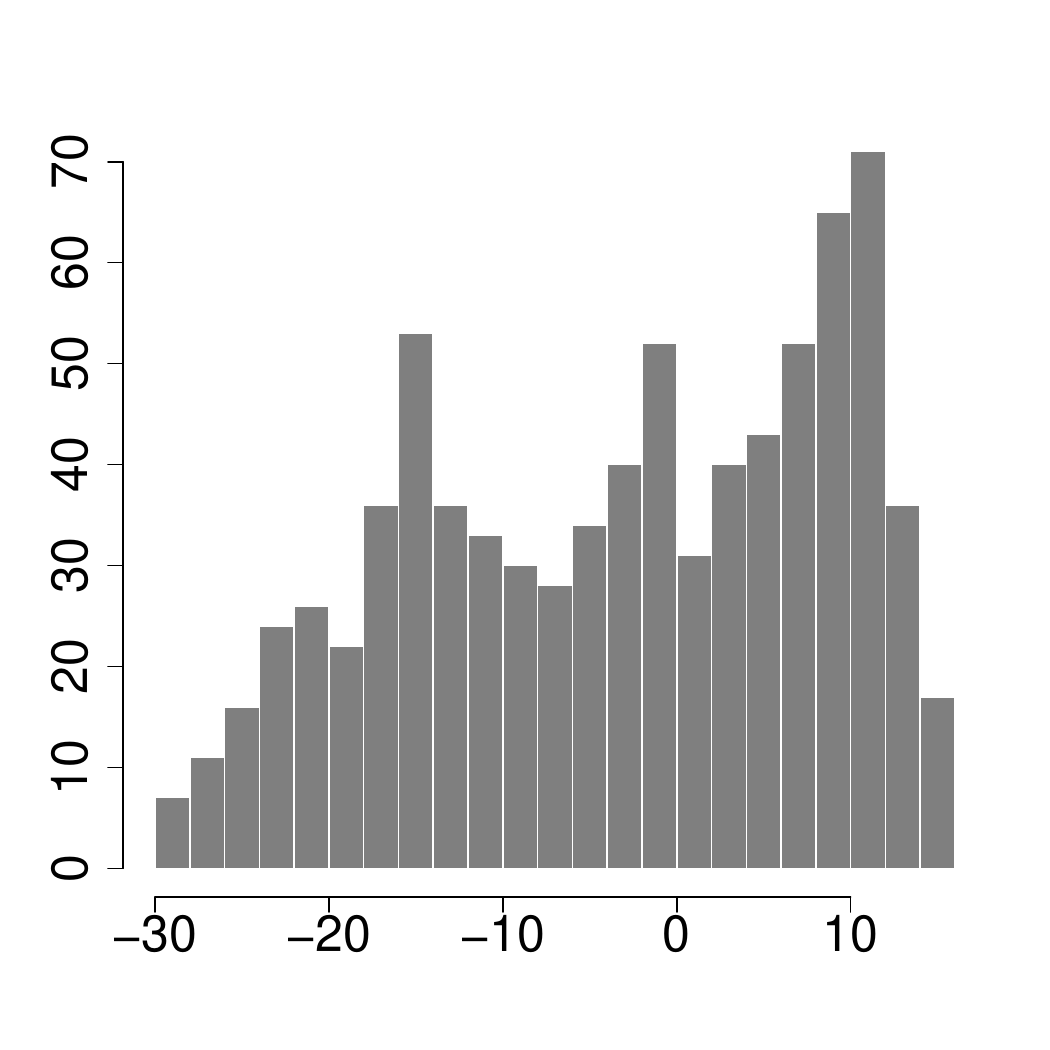} &
    \includegraphics[width=.15\textwidth]{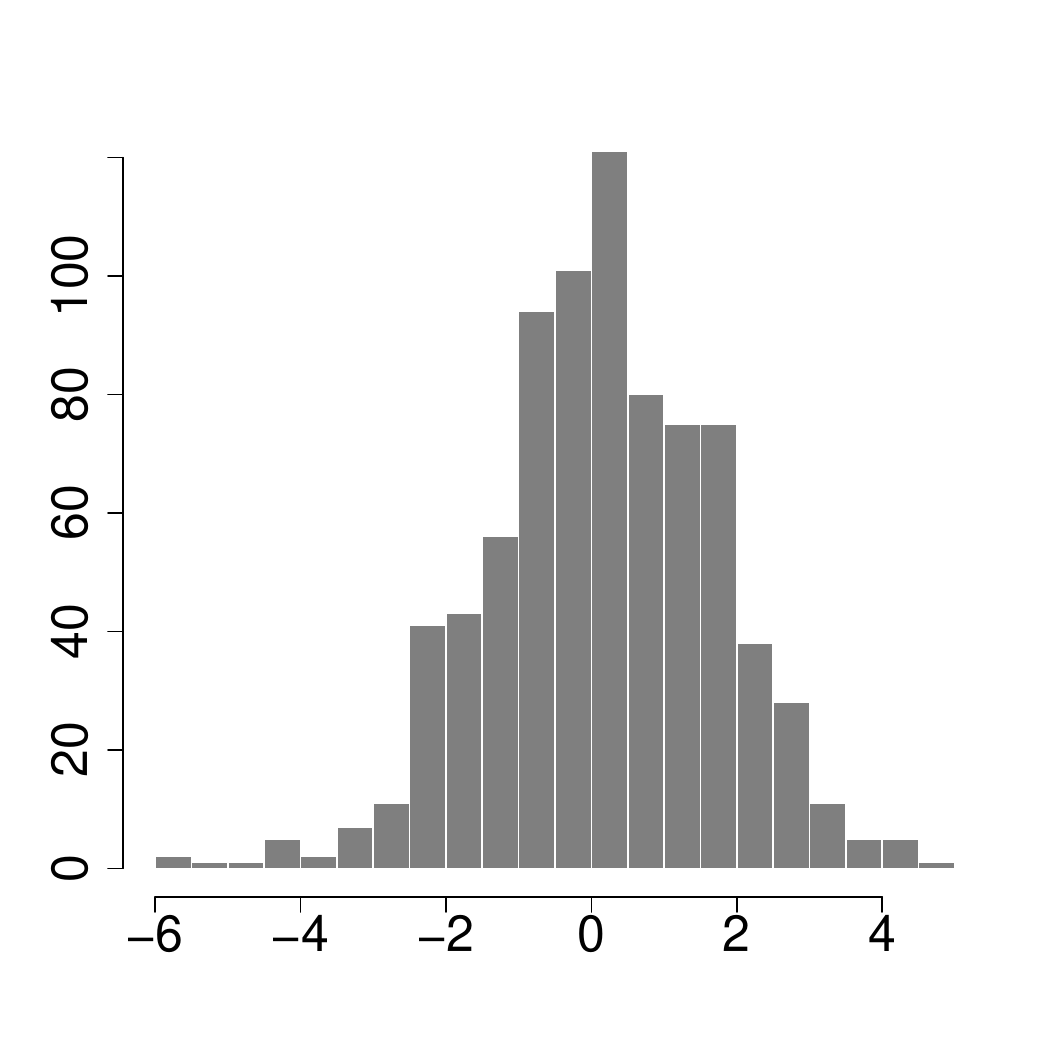} &
    \includegraphics[width=.15\textwidth]{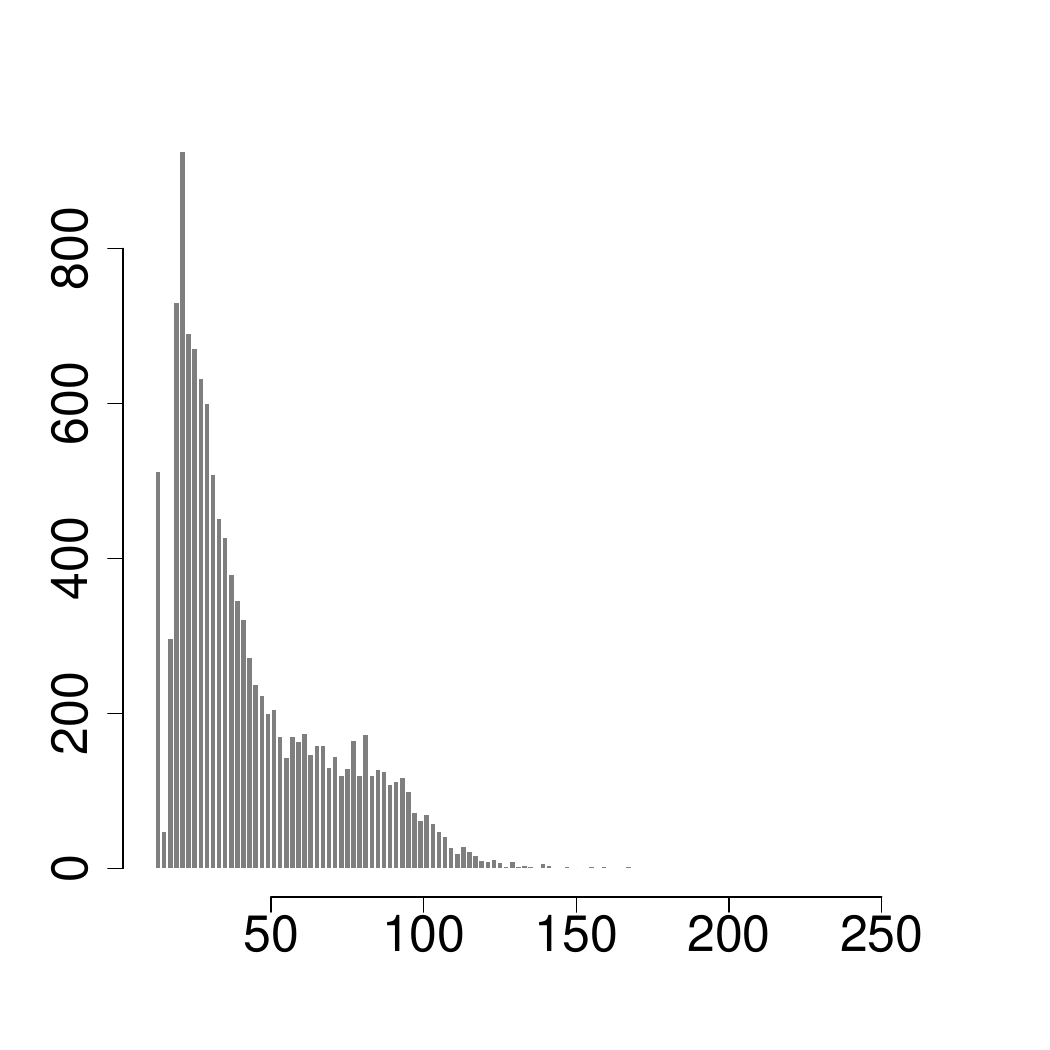} &
    \includegraphics[width=.15\textwidth]{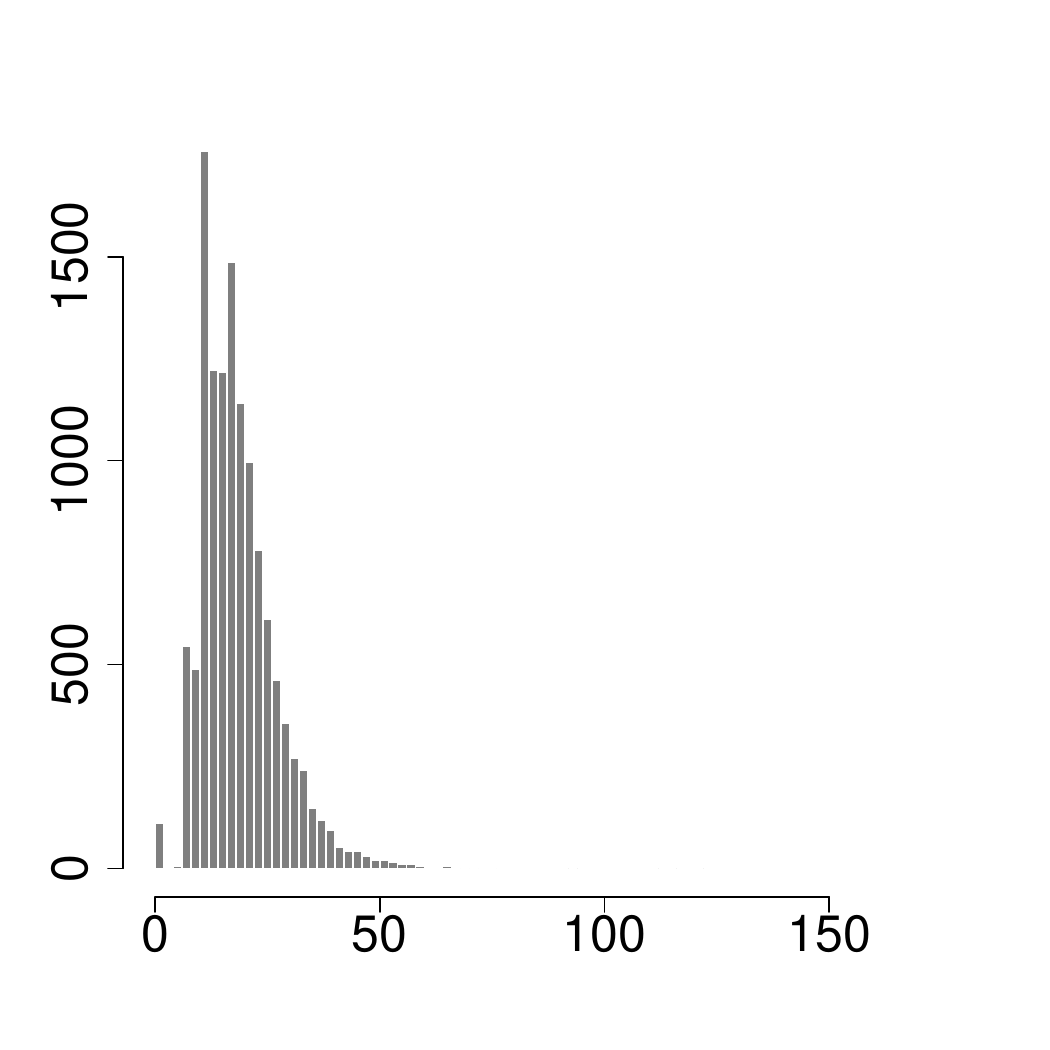} &
    \includegraphics[width=.15\textwidth]{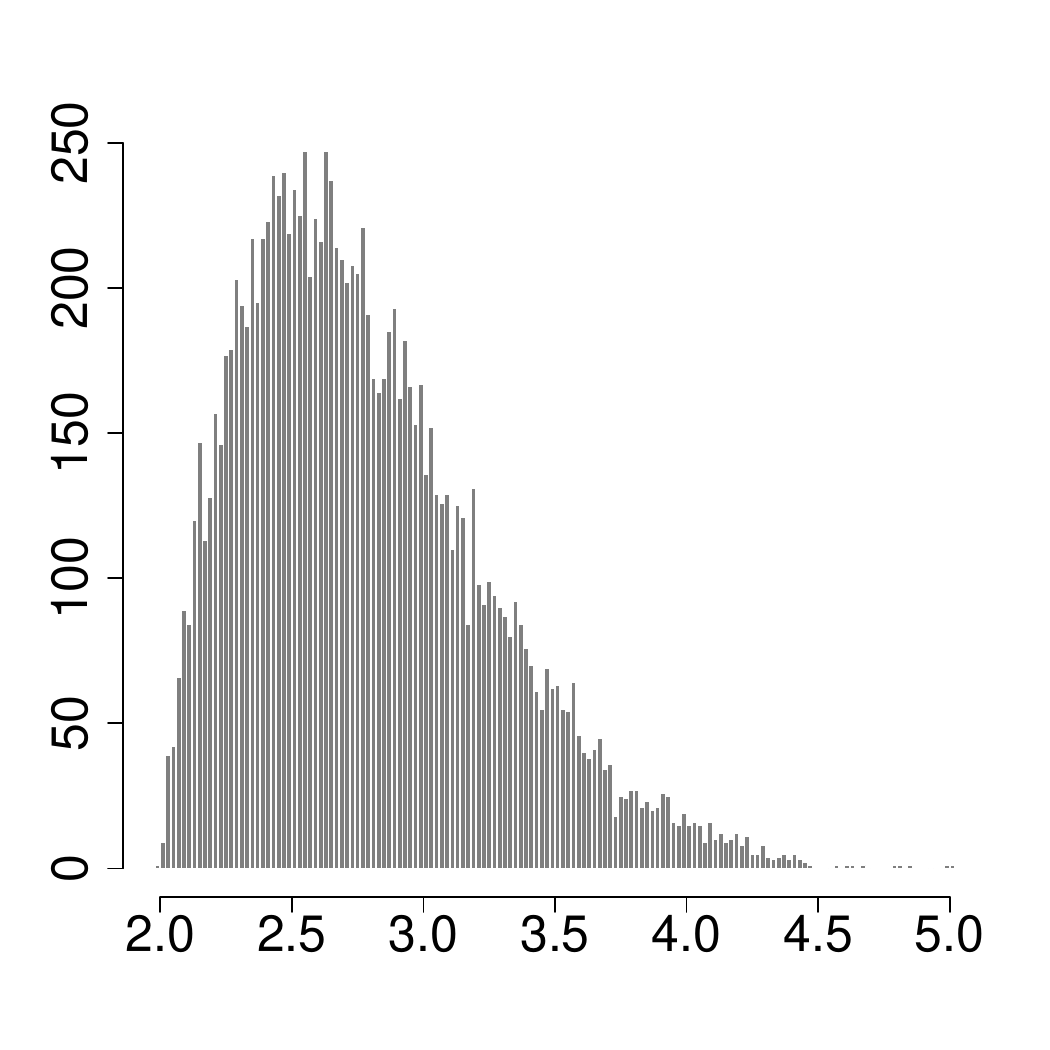} \\
    \includegraphics[width=.15\textwidth]{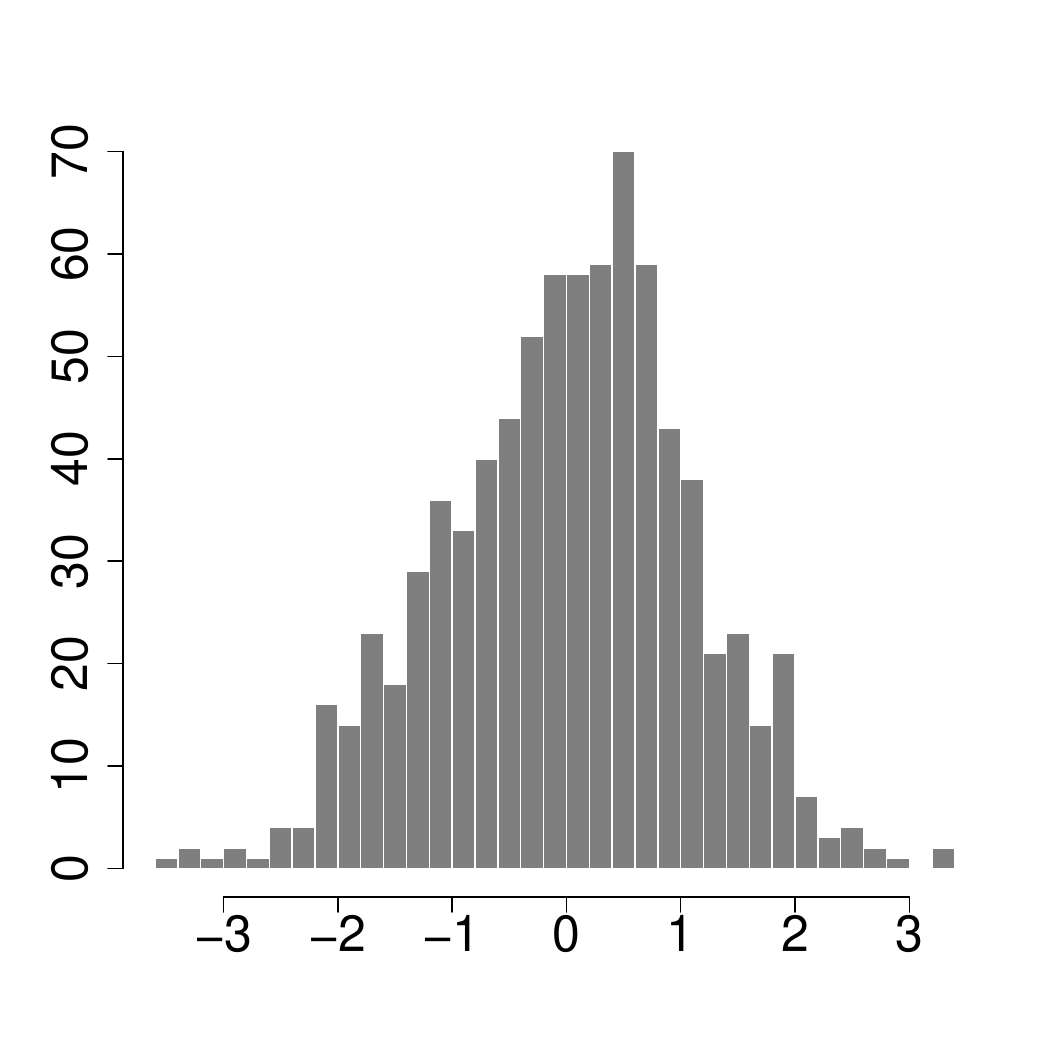} &
    \includegraphics[width=.15\textwidth]{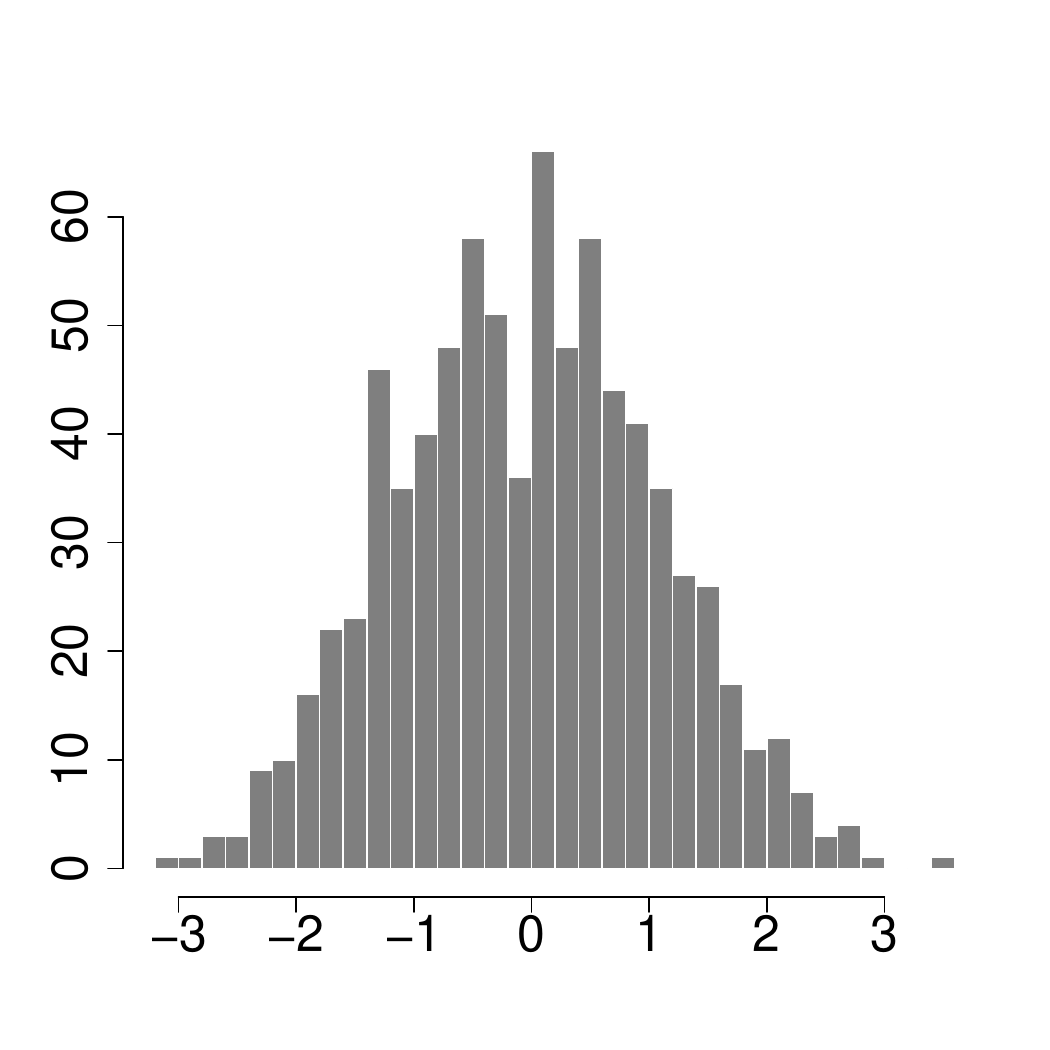} &
    \includegraphics[width=.15\textwidth]{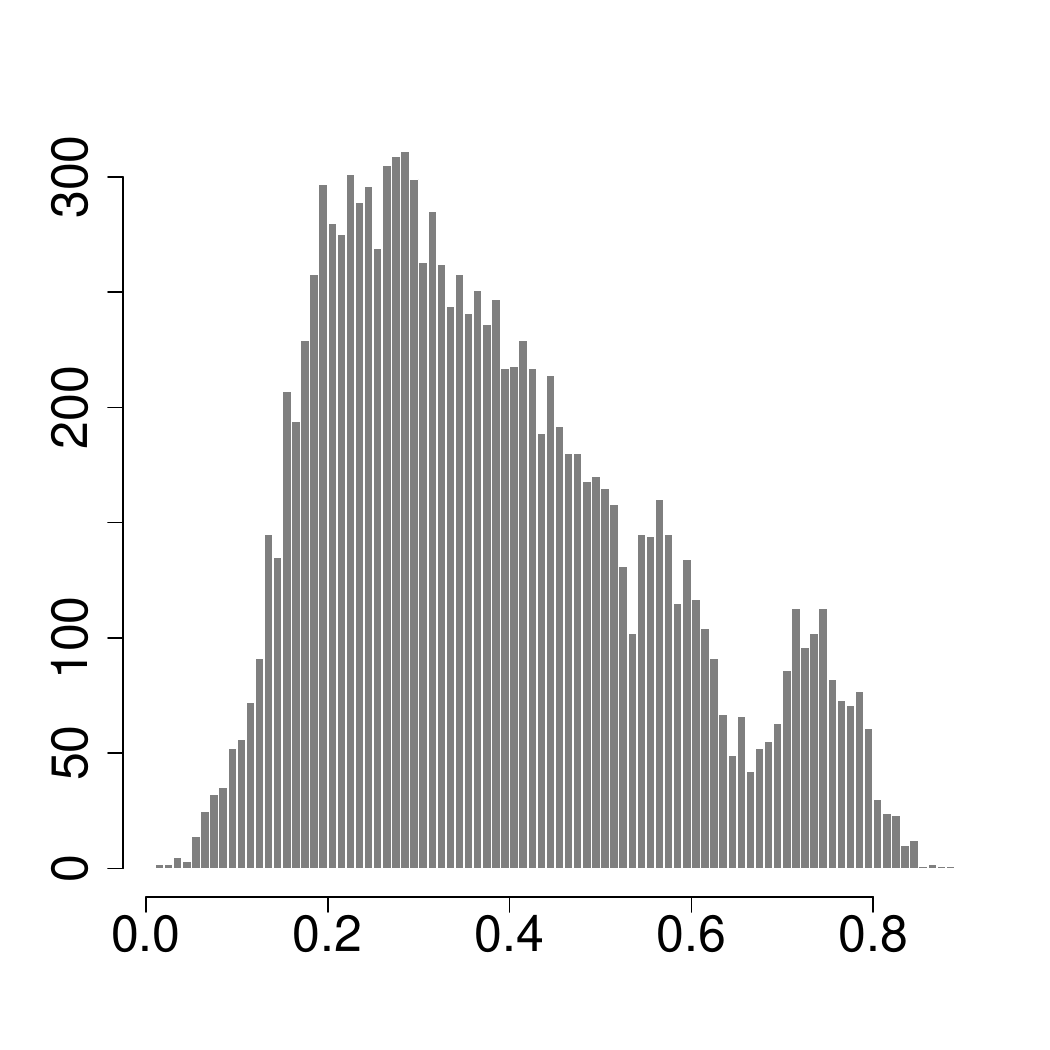} &
    \includegraphics[width=.15\textwidth]{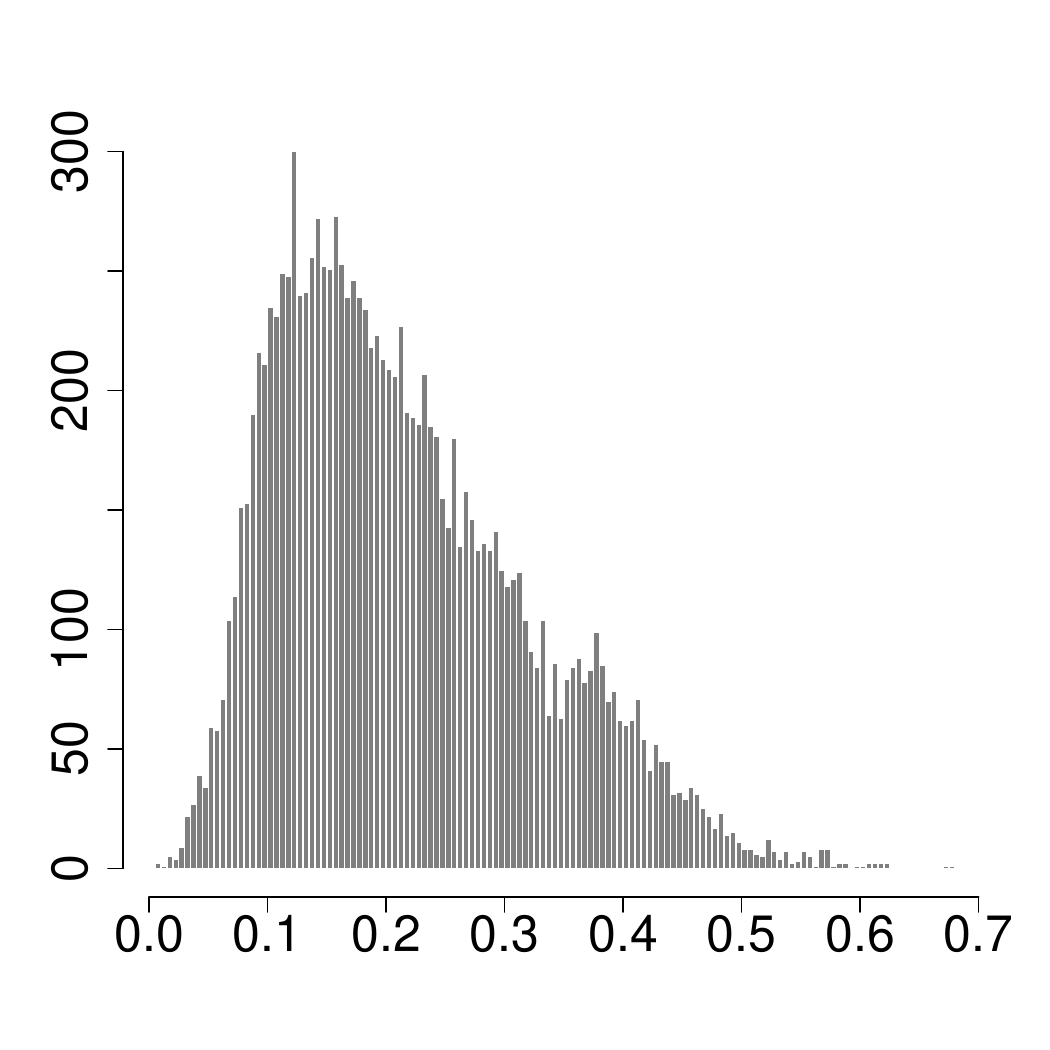} &
    \includegraphics[width=.15\textwidth]{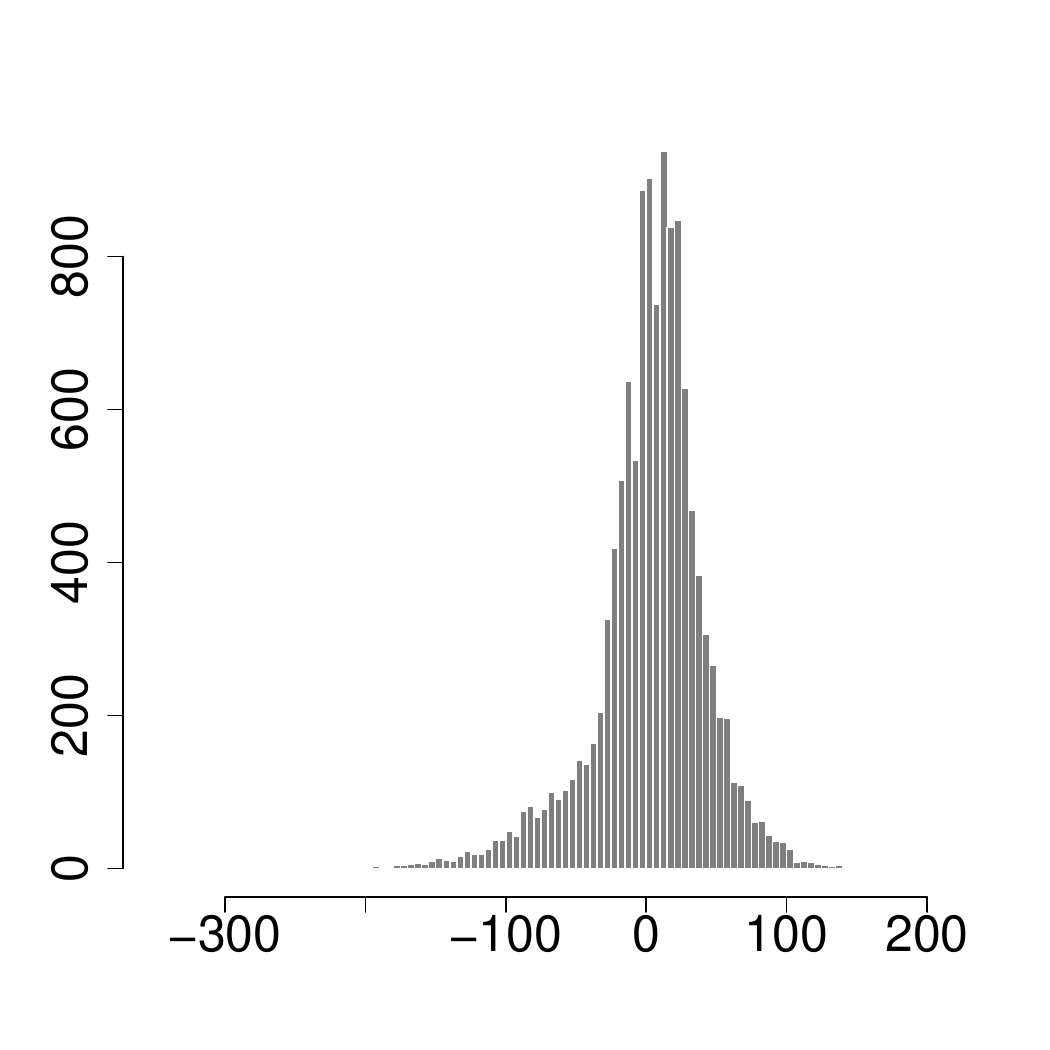} \\
    \includegraphics[width=.15\textwidth]{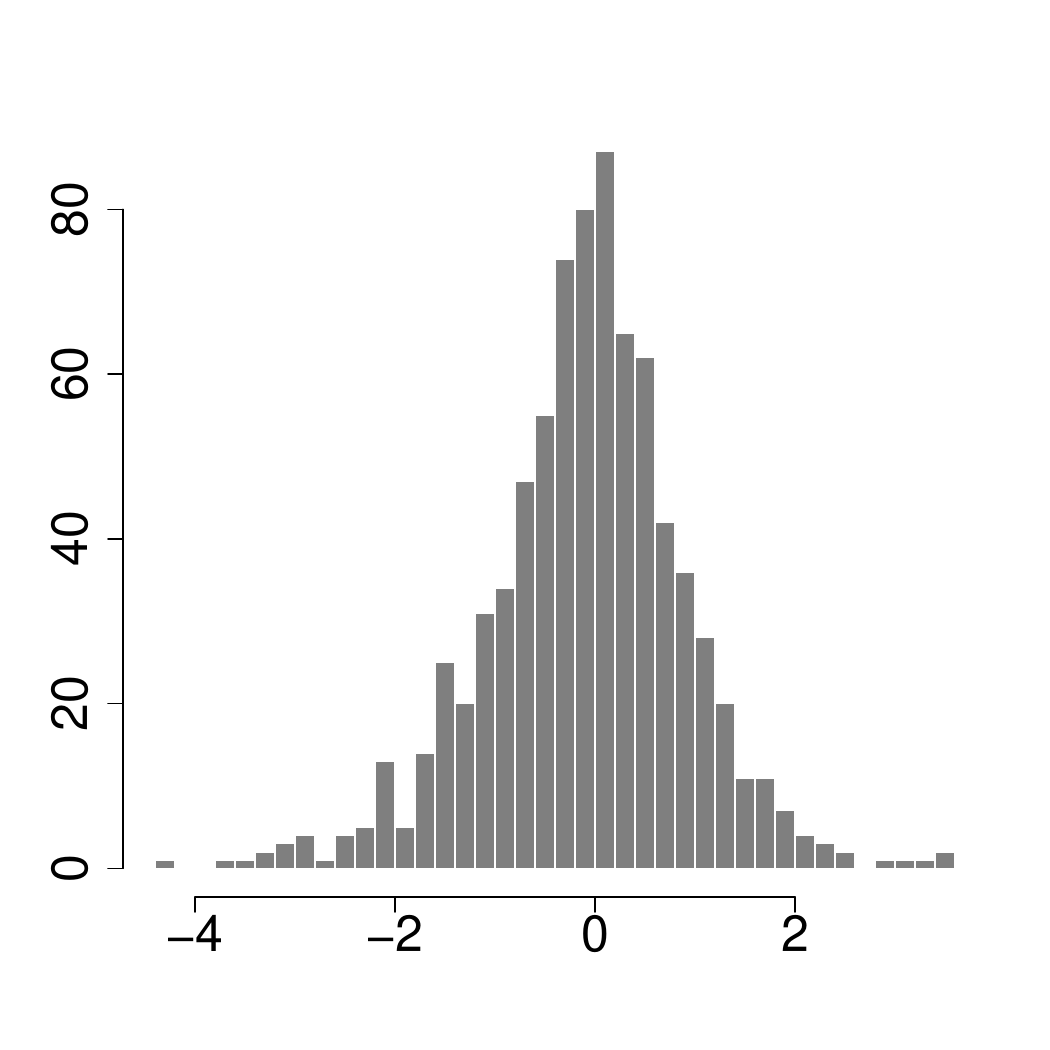} & 
    \includegraphics[width=.15\textwidth]{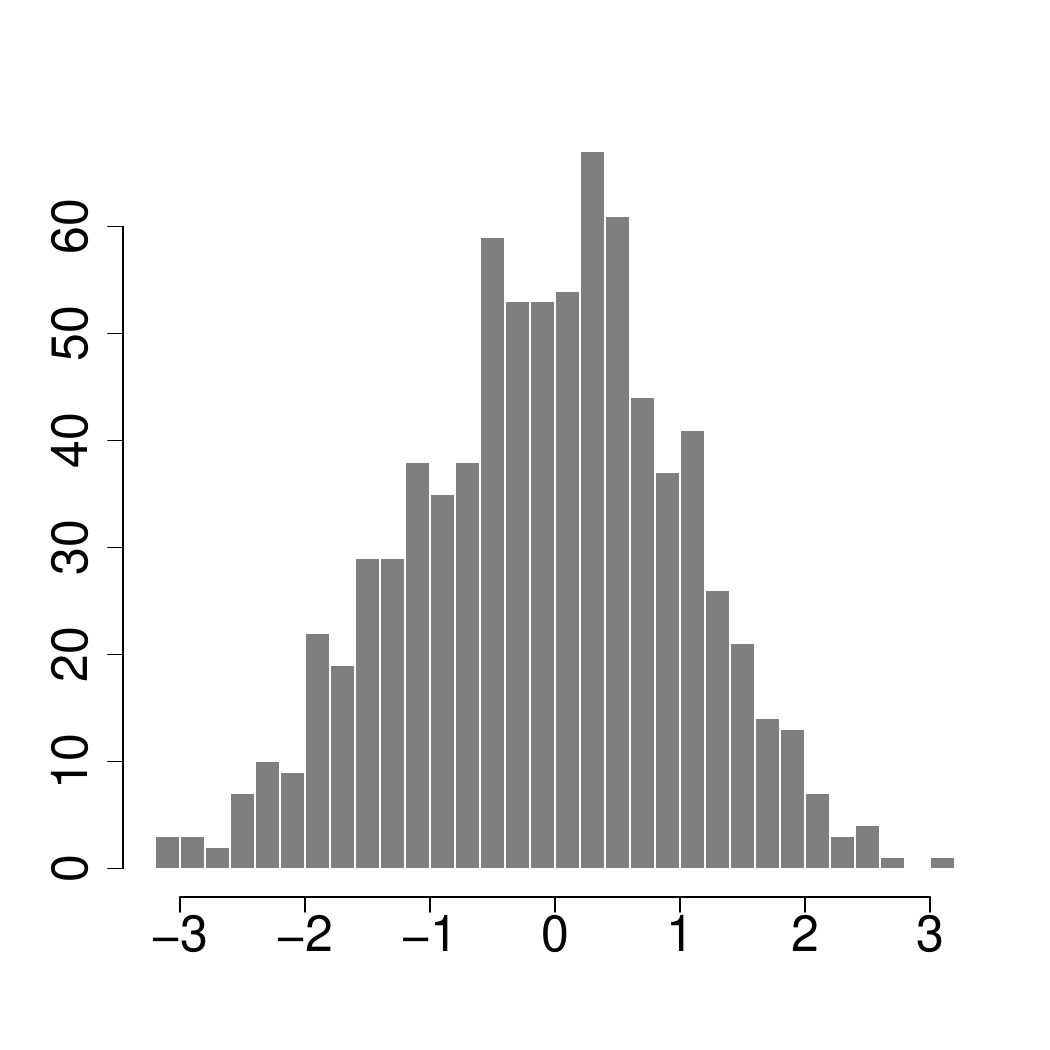} &
    \includegraphics[width=.15\textwidth]{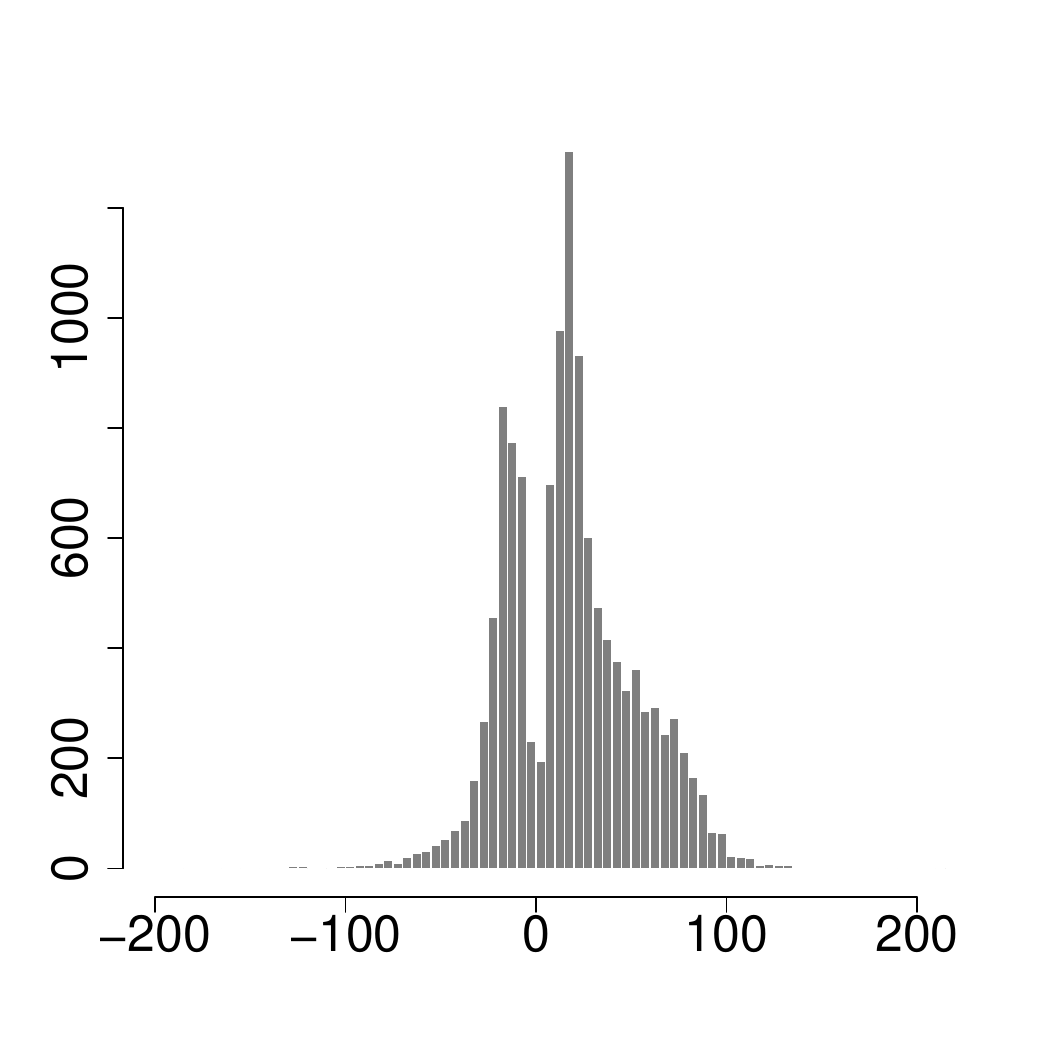} &
    \includegraphics[width=.15\textwidth]{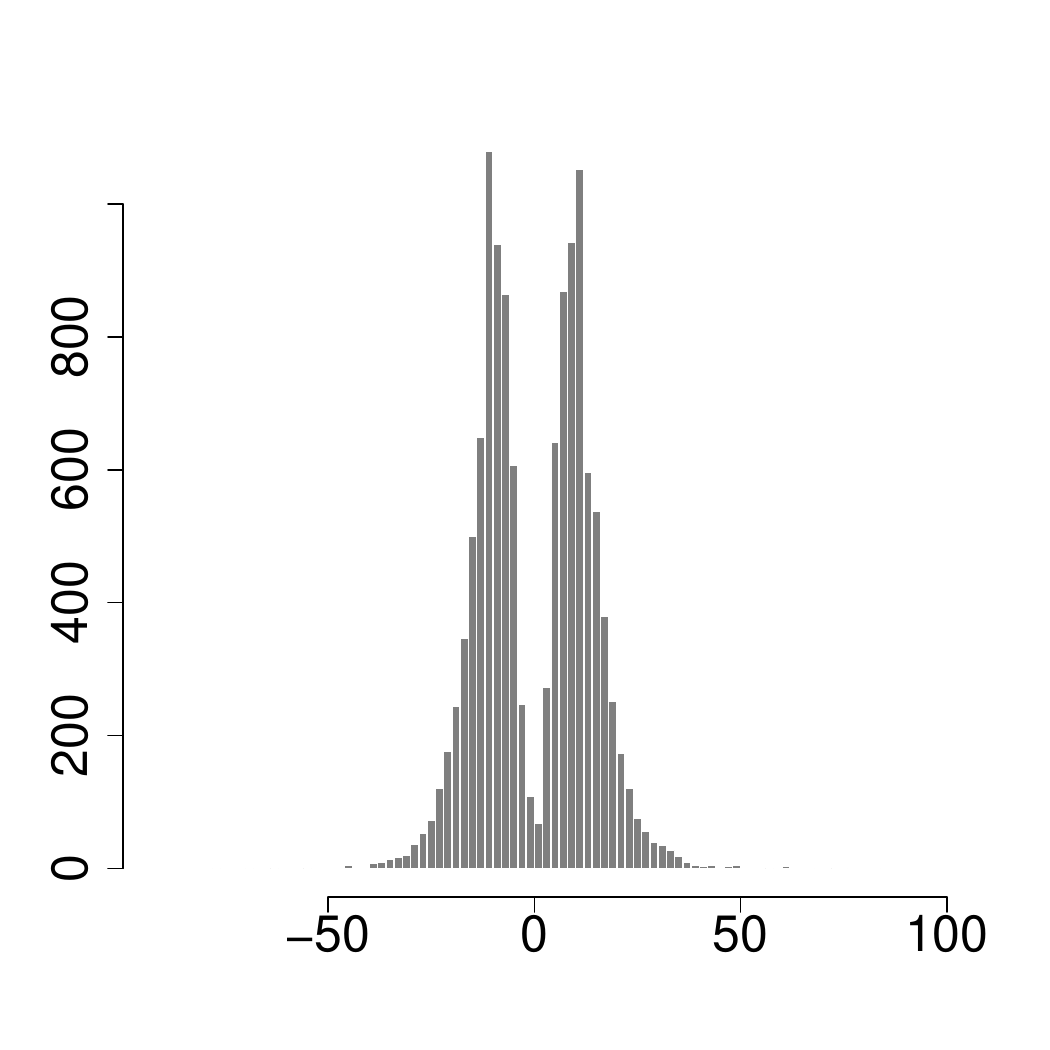} &
    \includegraphics[width=.15\textwidth]{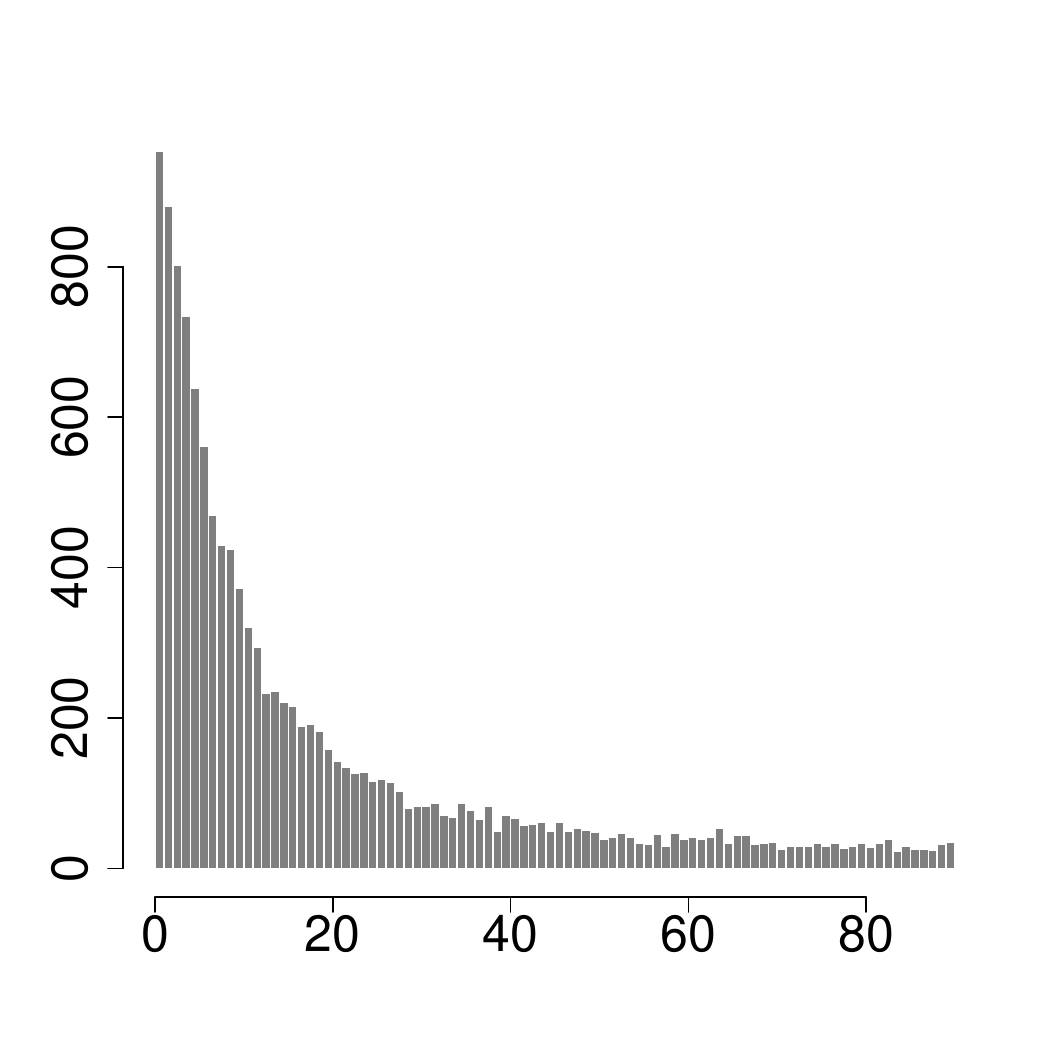} \\
    & & \includegraphics[width=.15\textwidth]{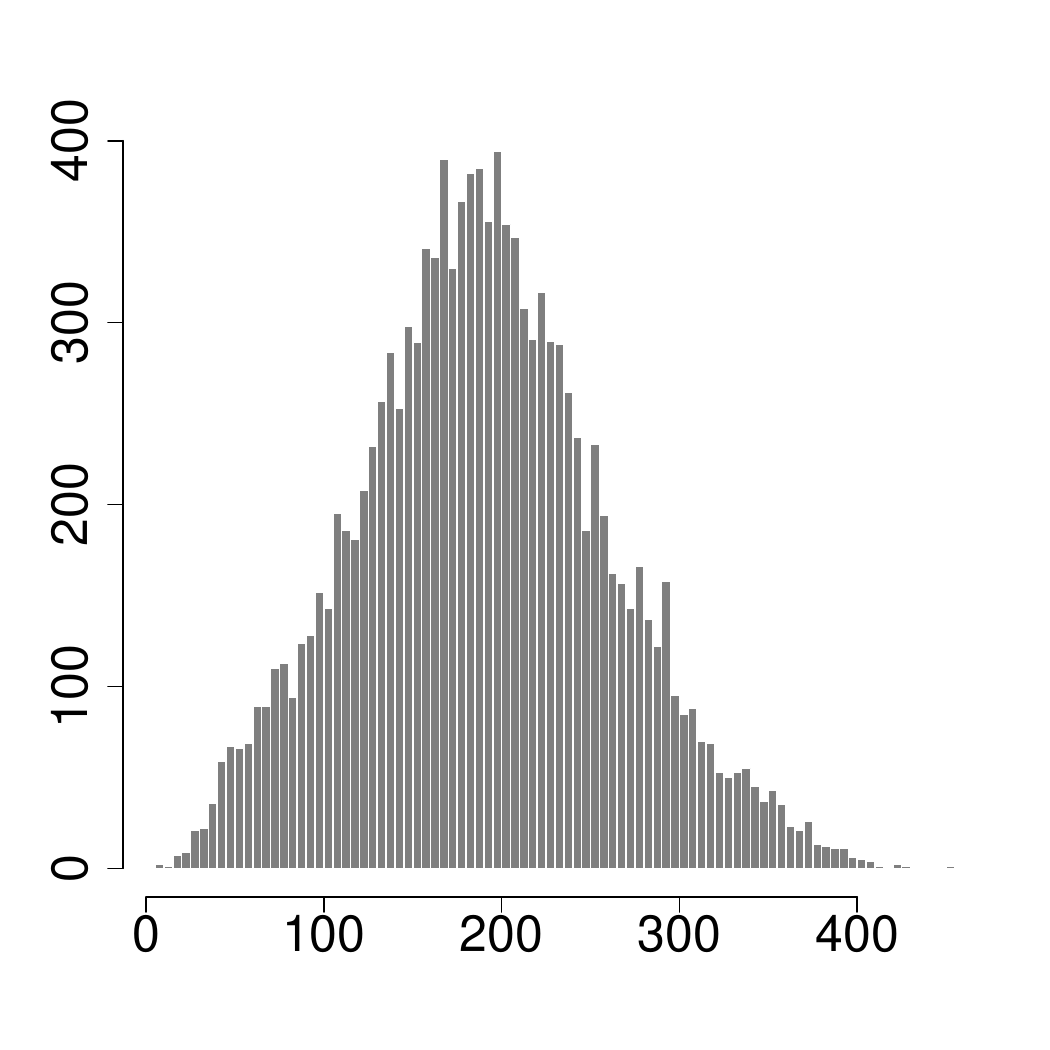} 
  \end{tabular}
  \caption{\small Histograms of the six climate indices (left) and ten telescope components (right).
  }
  \label{fig:hist}
\end{figure}

Gaussian graphical models are among the most popular because all quantities relevant to infer a graphical model can be easily estimated. Still, in many applications, either the observed variables are not marginally Gaussian or their joint distribution is far from being normal (see Figure \ref{fig:hist}).
We illustrate here how the entropy estimates we propose can be used to infer tree-shaped graphical models when dealing with continuous but non Gaussian data.

We consider two datasets. The first one was first introduced by \cite{ChG08} and further analyzed by \cite{Til09}; it consists of six climate indices describing the behaviour of the main oceans at a global scale. The second one is the MAGIC Gamma Telescope dataset analysed by \cite{Kir07}, which consists of ten real-valued components corresponding to high energy gamma rays \citep[see][for a precise description]{DuG17}. The variables included in both datasets display either non Gaussian marginal (see Figure \ref{fig:hist}) or non Gaussian joint distributions 
(not shown).
For the telescope dataset, we used the bounded GMM proposed by \cite{Scr19} for all  variables except the sixth, seventh, eighth and tenth.

For both datasets, understanding the dependency structure that relates all variables provides insights about the behavior of the global systems (i.e. ocean climate or gamma rays spectrum). To this aim, we determined the maximum-likelihood (tree-shaped) graphical model using \cite{ChL68}'s algorithm. More specifically, we estimated the mutual information between each pair of variables of each dataset either ($a$) doing as if all variables were Gaussian or ($b$) using the estimate we propose. 

\begin{figure}[htb]
  \centering
  \begin{tabular}{cc|cc}
    \multicolumn{2}{c|}{Climate dataset} & \multicolumn{2}{c}{Telescope dataset} \\
    Gaussian & Gaussian mixture & Gaussian & Bounded Gaussian mixture \\
    \hline
    \includegraphics[width=.22\textwidth]{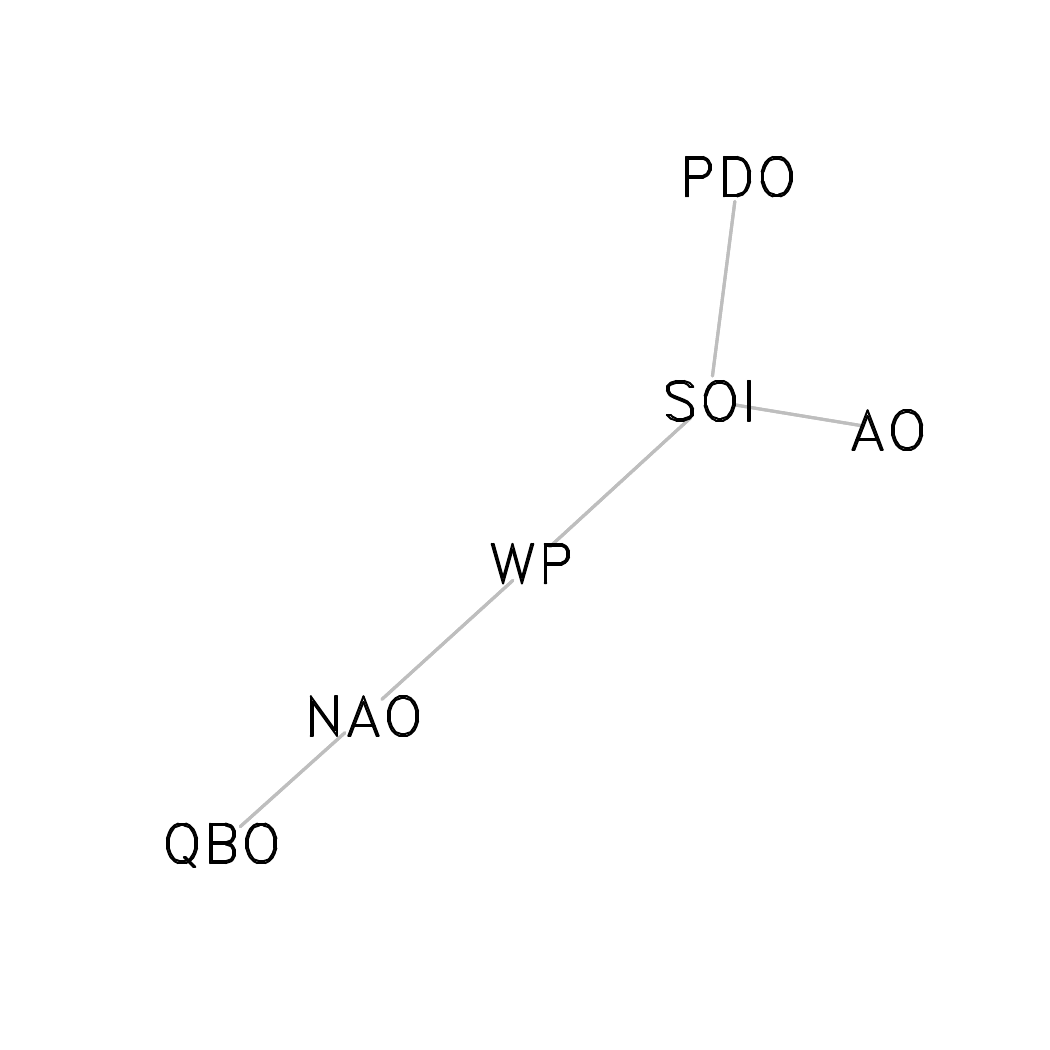} &
    \includegraphics[width=.22\textwidth]{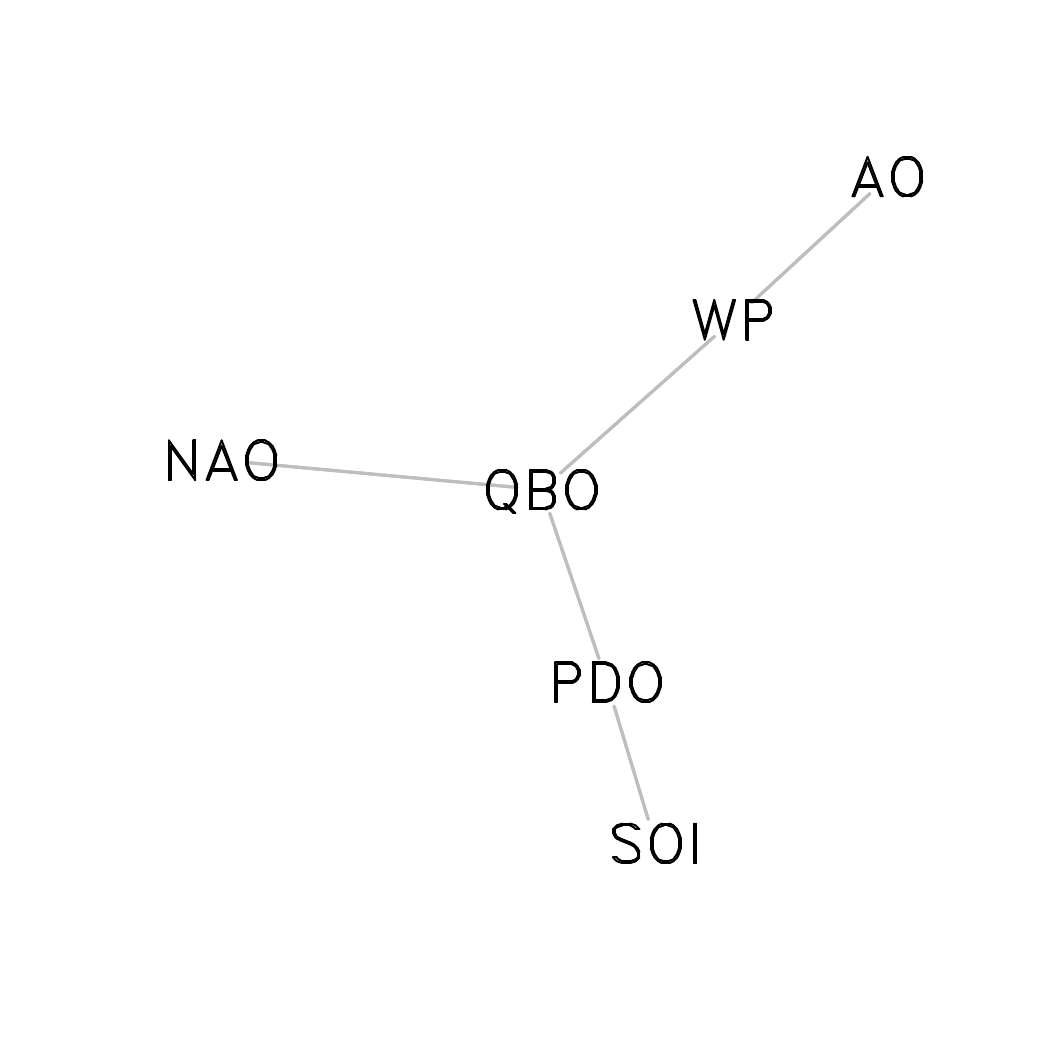} &
    \includegraphics[width=.22\textwidth]{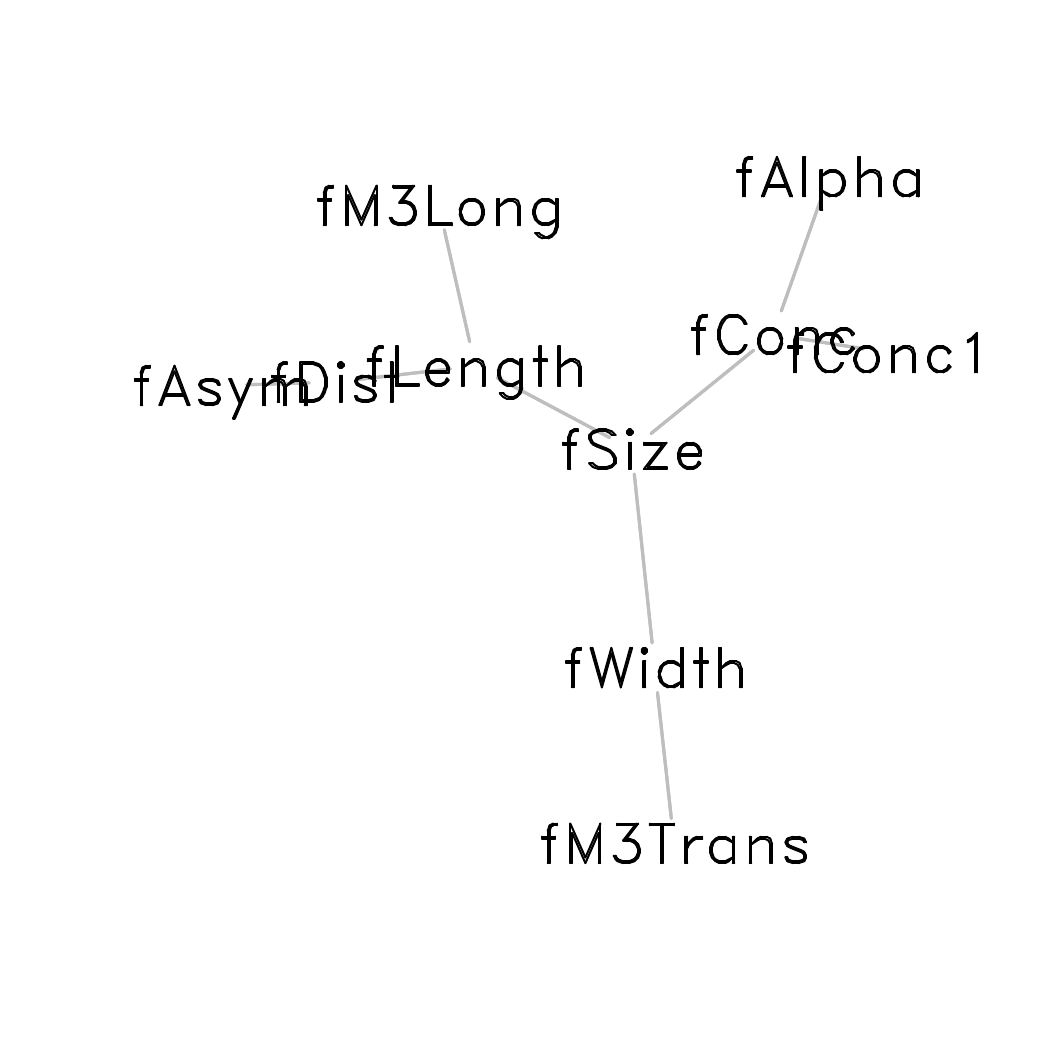} &
    \includegraphics[width=.22\textwidth]{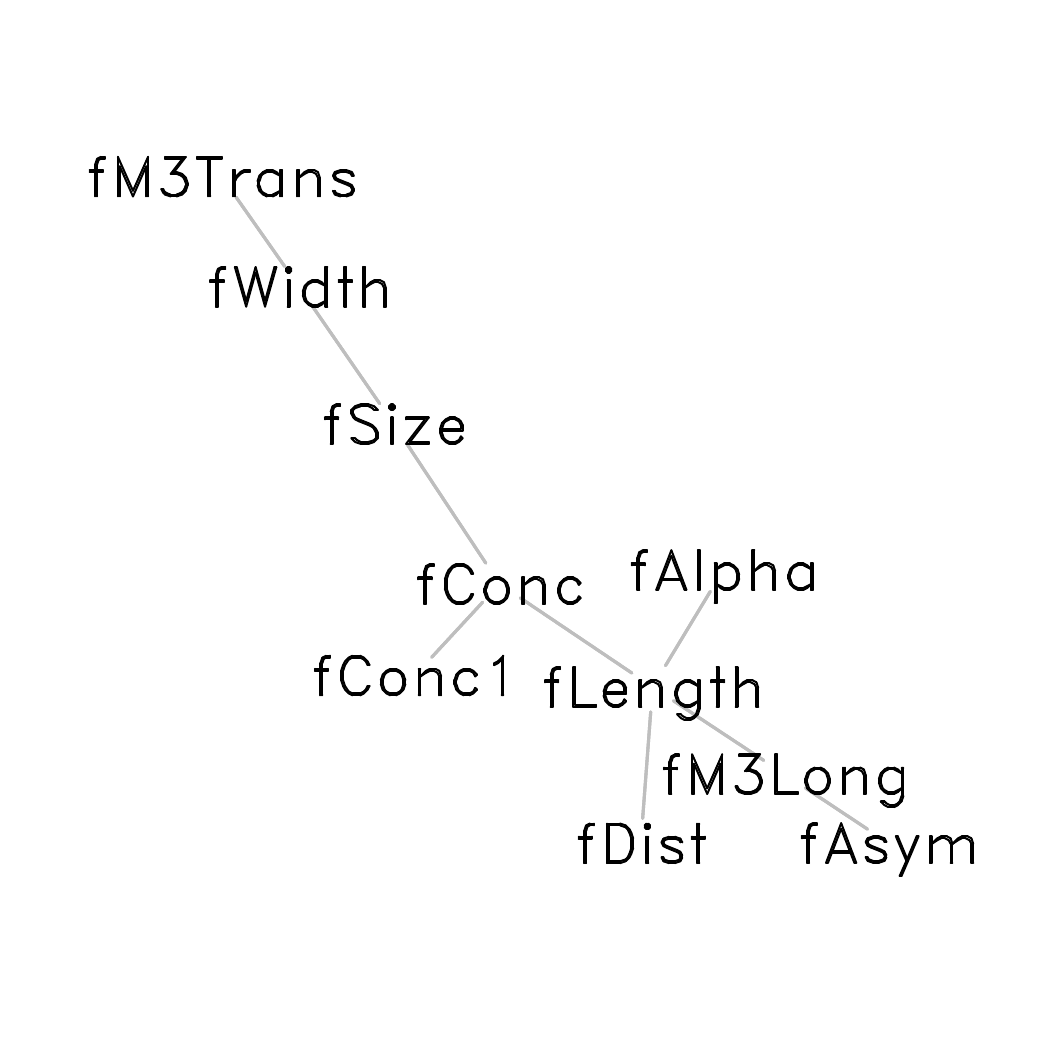} \\
    \hline
    \includegraphics[width=.22\textwidth]{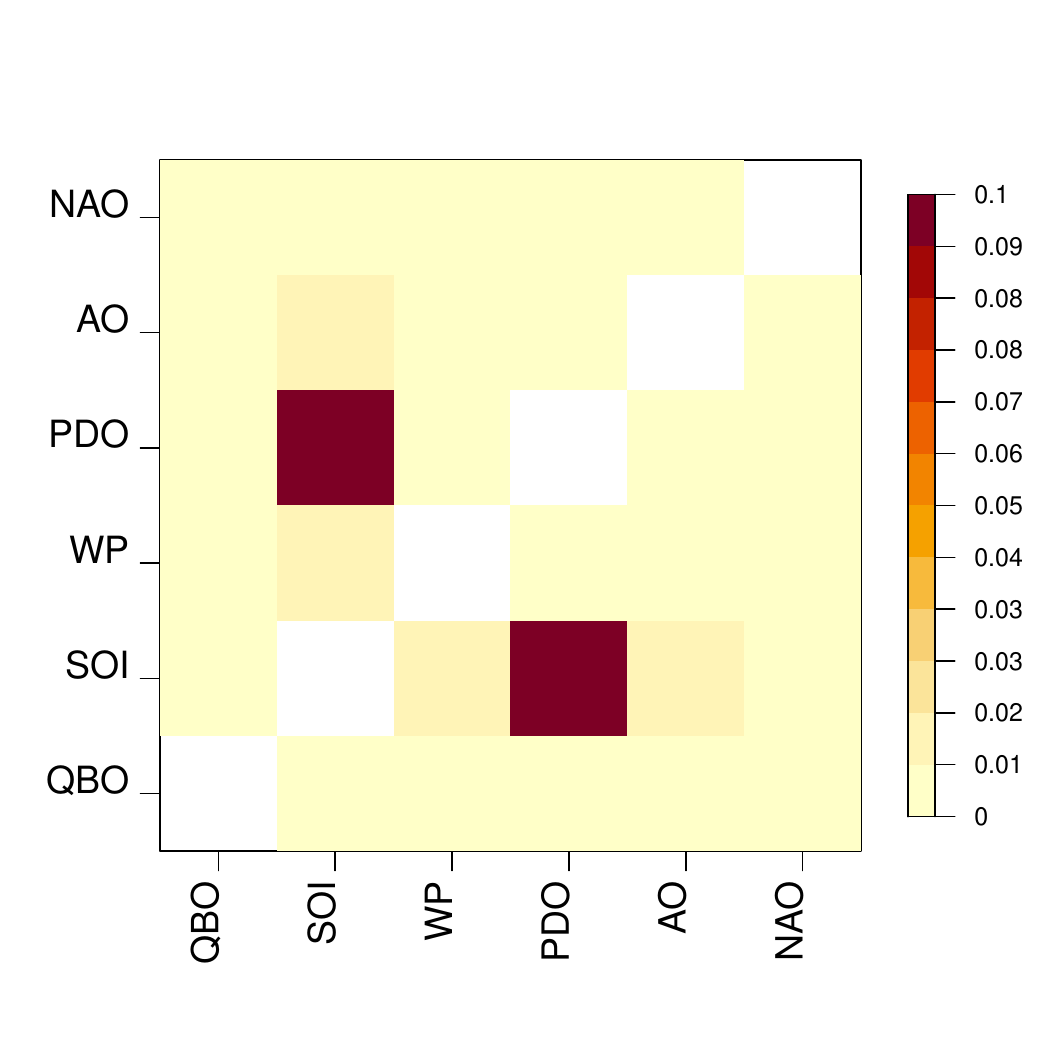} &
    \includegraphics[width=.22\textwidth]{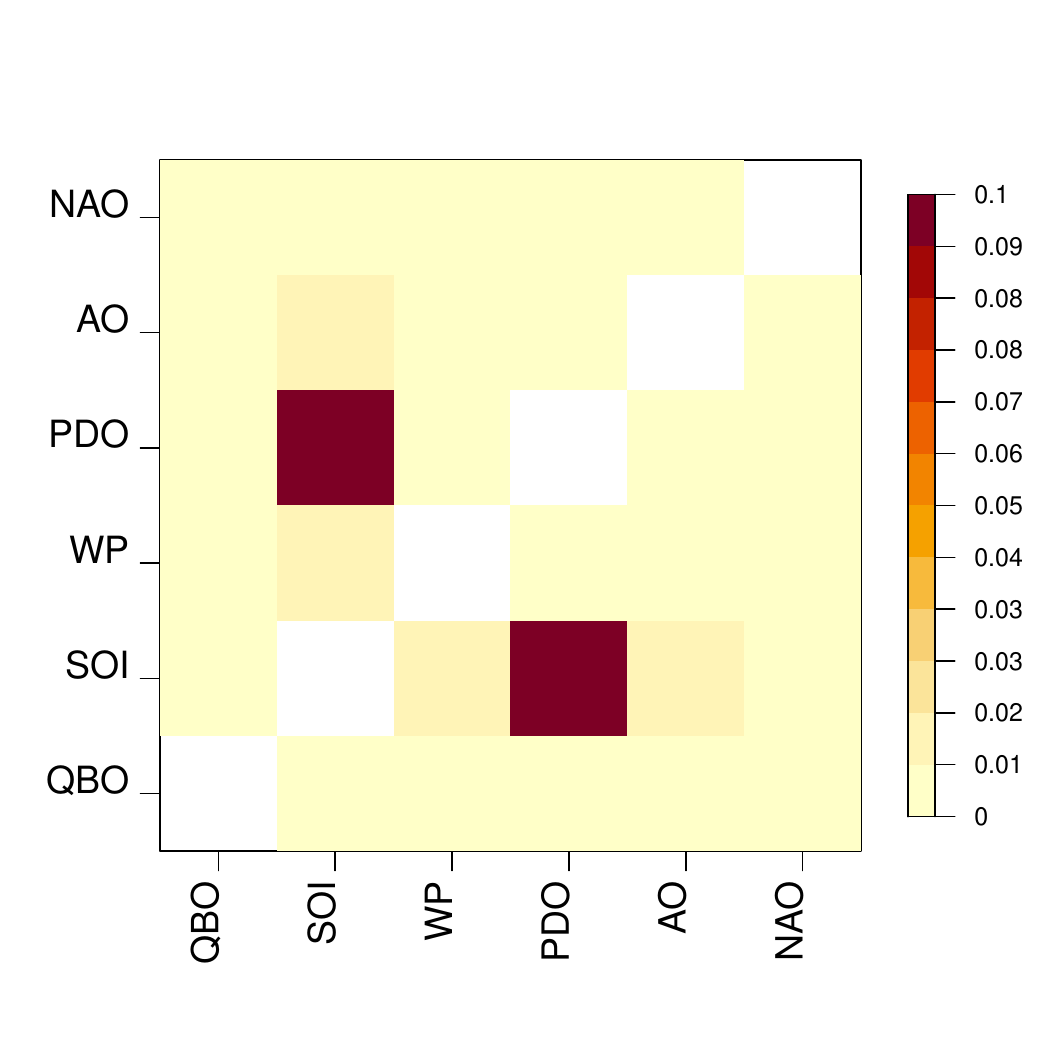} &
    \includegraphics[width=.22\textwidth]{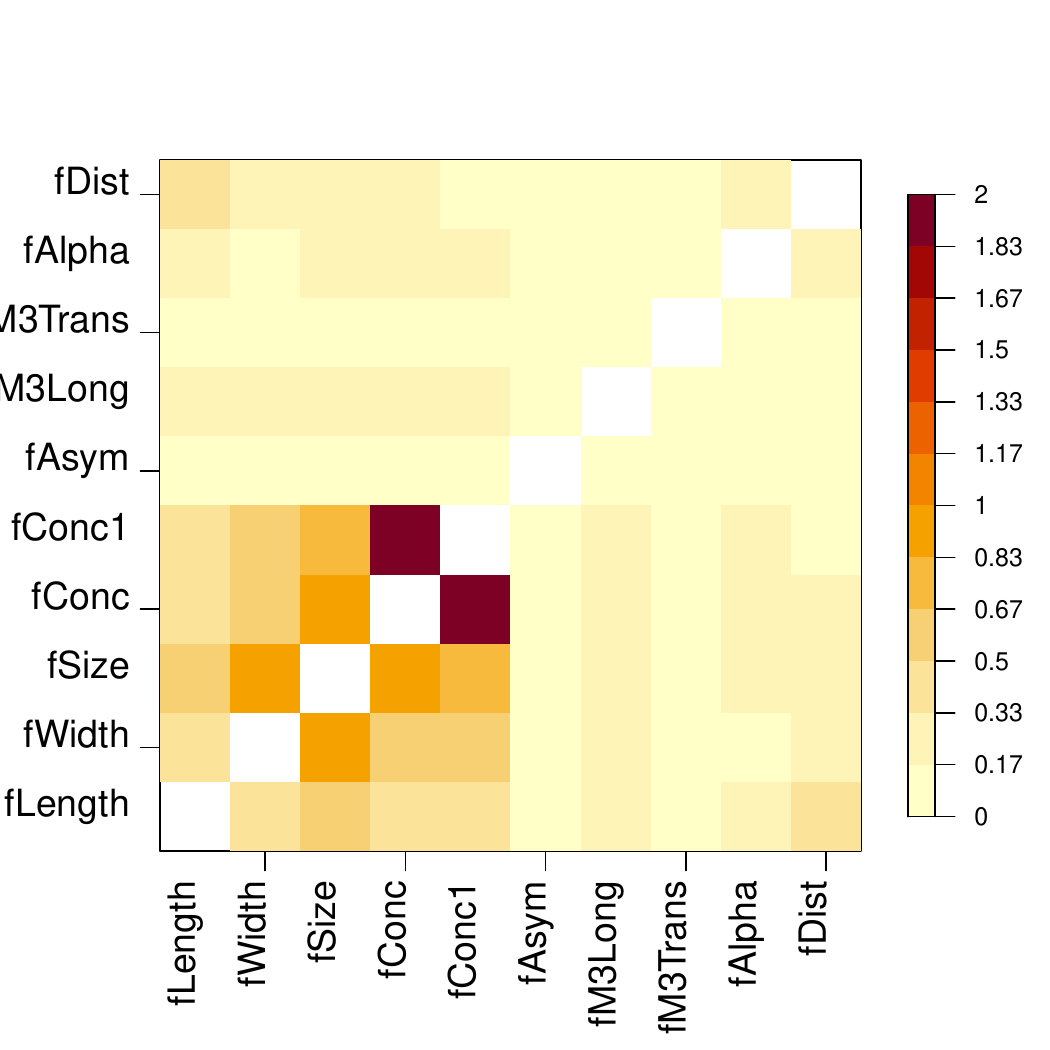} &
    \includegraphics[width=.22\textwidth]{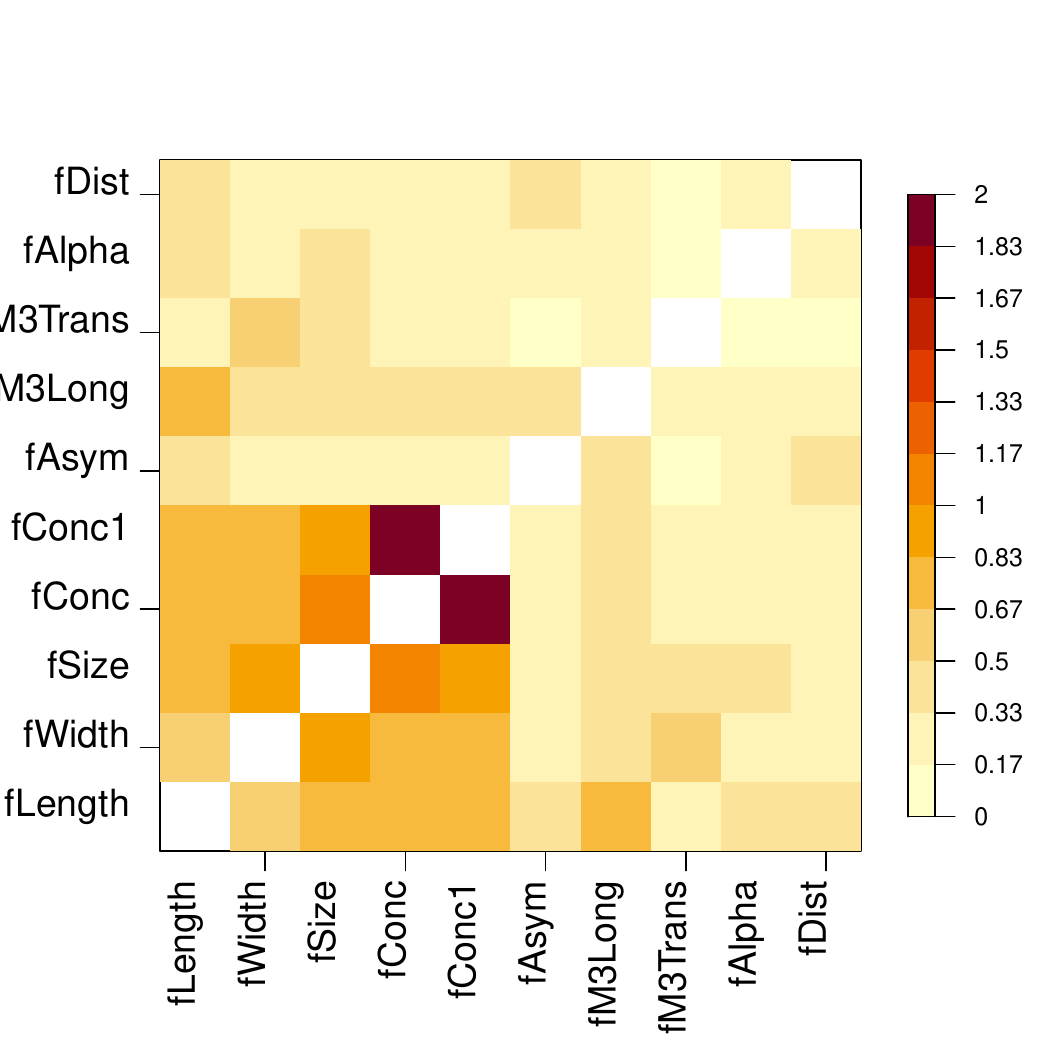} \\
  \end{tabular}
  \caption{\small Maximum likelihood trees (top panels) and estimated mutual informations (bottom panels) for the climate and telescope datasets.
  } 
  \label{fig:treeGM}
\end{figure}
      
Figure \ref{fig:treeGM} (top) shows that, for both datasets, the inferred graphical models are considerably modified when using a naive Gaussian model, as compared to the estimation we propose, which accounts for the non-normality of the data. For the climate dataset, only 2 edges (among 5) are common to the Gaussian and the GMM trees. The effect is less drastic for the telescope data, where 6 edges (among 9) are common to the Gaussian and the (bounded) GMM trees. Not accounting for the boundedness of the variables also has a strong impact as only 5 edges are common to the regular GMM and to the bounded GMM trees (not shown). Indeed, no ground truth is available for either of theses examples. Still, Figure \ref{fig:treeGM} (bottom) shows that the normality assumption yields in strong smoothing of the estimated mutual information, making the graphical model inference harder, and therefore less reliable.

\clearpage

\section{Conclusion} \label{sec:conc}
In this paper, we proposed a versatile estimate of the entropy and related quantities for a large class of multivariate distributions. We illustrated the 
{role this estimate can have in various problems}, including image processing and network inference. 
{The accuracy and efficiency of the proposal are demonstrated through extensive simulation results}. Importantly, this estimate is easy to implement using any computationally efficient package fitting mixtures of multivariate distributions, such as \texttt{mclust} \citep{Mclust} or \texttt{flexmix} \citep{FlexMix}, to name a few.

The proposed methodology opens a series of interesting questions. 
For example, rather than using only the entropy estimate provided by the the mixture model with best BIC, the estimates provided by mixtures with different number of components could be combined using Bayesian model averaging \citep[BMA:][]{HMR99}, with no additional computational cost. 
Finally, although we showed that the proposed estimator compares well with alternative proposals from an empirical view-point, its sampling behaviour and properties should be investigated from a theoretical point of view to evaluate its consistency and to derive its (asymptotic) variance.


\appendix
\section{Approximations for the entropy of a GMM} 
\label{sec:approxEntropyGMM}
Several approximations to the entropy of a random variable $Y$ with Gaussian mixture distribution $f(y) = \sum_{k=1}^K \pi_k \phi(y; \mub_k, \Sigmab_k)$ have been proposed in the literature. 
{In the following, we briefly describe some approximations that have been used in the simulation studies in Section~\ref{sec:sim}.}

\begin{description}
\item[Unscented Transformation (UT):]
A first approximation can be obtained using the Unscented Transformation approach \citep{Julier:Uhlmann:1996, Goldberger:Aronowitz:2005}, where the required integral {in \eqref{eq:entropy}} is approximated by computing an average of the log-density evaluated over the set of $2p$ so-called \emph{sigma-points}. Differently from MC integration, such points are chosen deterministically. 
Let the sigma-points be defined as 
\begin{equation*}
\left\{
\begin{aligned}
\widetilde{y}_{k,j} & = \mub_k + \left[\sqrt{p\Sigmab_k}\right]_j
&&& j = 1, \dots, p \\
\widetilde{y}_{k,(p+j)} & = \mub_k - \left[\sqrt{p\Sigmab_k}\right]_j
&&& j = 1, \dots, p \\
\end{aligned}
\right.
\end{equation*}
where $\left[\sqrt{p\Sigmab_k}\right]_j$ is the $j$th column of the square root matrix of $p\Sigmab_k$, so $\left[\sqrt{p\Sigmab_k}\right]_j = \sqrt{p\lambda_j}\u _j$ with $\lambda_j$ and $\u _j$ being, respectively, the $j$th eigenvalue and eigenvector of $\Sigmab_k$. 
Then, the UT entropy approximation is computed as
\begin{equation*}
\Entropy_{\text{UT}}(Y) = -\frac{1}{2p} \sum_{k=1}^K \pi_k \sum_{j=1}^{2p} \log{f(\widetilde{y}_{k,j})}.
\end{equation*}
\item[Variational approximation (VAR):]
A variational approximation for the entropy can also be obtained {\citep{Hershey:Olsen:2007} by} computing
\begin{align*}
  \Entropy_{\text{VAR}}(Y) 
  & = \sum_{k = 1}^{K} \pi_k \log \sum_{\ell=1}^{K} \pi_\ell 
  \exp\{KL\left( \mathcal N(\mub_k, \Sigmab_k) || \mathcal N(\mub_\ell, \Sigmab_\ell) \right)\} \\
  & \quad - \sum_{k=1}^{K} \pi_k H\left(\mathcal N(\mub_k, \Sigmab_k) \right),
\end{align*}
where $KL\left(\mathcal N(\mub_k, \Sigmab_k) || \mathcal N(\mub_\ell, \Sigmab_\ell) \right)$ is the Kullback-Leibler divergence between two Gaussian random variables with respective parameters $(\mub_k, \Sigmab_k)$ and $(\mub_\ell, \Sigmab_\ell)$, and $H\left(\mathcal N(\mub_k, \Sigmab_k)\right)$ is the entropy of the $k$th Gaussian component of the mixture. 
Note that the KL-divergence between two multivariate Gaussian distributions also admits a closed-form expression \citep[see for example][]{CoA12}.
\item[Second order Taylor expansion (SOTE):]
\citet{Huber:etal:2008} proposed a \emph{second order Taylor expansion} to approximate the entropy {of} GMMs {by writing}
\begin{equation*}
\Entropy_{\text{SOTE}}(Y) = \Entropy_0(Y) - \sum_{k=1}^{k} \dfrac{\pi_k}{2} F(\mub_k) \odot \Sigmab_k,
\end{equation*}
where $\Entropy_0(Y) = -\sum_{k=1}^{K} \pi_k \log \phi(\mub_k ; \mub_k, \Sigmab_k)$ is the first order expansion of the entropy around the mean vector $\mub_k$, $\odot$ is the so-called matrix contradiction operator, so that for the two matrices $\A \in \mathbb{R}^{n \times m}$ and $\B \in \mathbb{R}^{n \times m}$, $\A \odot \B = \sum_{i=1}^{n}\sum_{j=1}^{m} a_{ij}b_{ij}$, and
\begin{align*}
F(x) & = \dfrac{1}{f(x)} \sum_{k=1}^K \pi_k \Sigmab^{-1}_k \left( \dfrac{1}{f(x)}(\mub_k - x) (\nabla f(x))\T + 
(\mub_k - x)\left(\Sigmab^{-1}_k (x - \mub_k)\right)\T - \I \right) \\
& \qquad \times \phi(x ; \mub_k, \Sigmab_k), 
\end{align*}
for a generic vector $x$, where $\nabla f(x)$ is the gradient of the mixture model with respect to the data.
\end{description}
Methods UT, VAR, and SOTE, have been recently used by \citet{Scrucca:Serafini:2019} for deriving a projection pursuit method aimed at maximizing the negentropy for Gaussian mixtures. Moreover, they are implemented in the \pkg{ppgmmga} R package \citep{pkg:ppgmmga}.

\bibliographystyle{authordate3} 
\bibliography{biblio}

\end{document}